\documentclass[12pt,a4paper]{article}
\pdfoutput=1

\usepackage{graphicx,psfrag}
\usepackage{bm}
\usepackage{mathbbol,verbatim}
\usepackage{slashed}
\usepackage{graphics}
\usepackage{color,ulem, amsmath, amssymb}
\usepackage{tikz}
\usetikzlibrary{arrows,shapes}
\usetikzlibrary{trees}
\usetikzlibrary{matrix,arrows} 				
\usetikzlibrary{positioning}				
\usetikzlibrary{calc,through}				
\usetikzlibrary{decorations.pathreplacing}  
\usepackage{pgffor}							

\usetikzlibrary{decorations.pathmorphing}	
\usetikzlibrary{decorations.markings}
\tikzset{
    vector/.style={decorate, decoration={snake}, draw},
	provector/.style={decorate, decoration={snake,amplitude=2.5pt}, draw},
	antivector/.style={decorate, decoration={snake,amplitude=-2.5pt}, draw},
    fermion/.style={draw=black, postaction={decorate},
        decoration={markings,mark=at position .55 with {\arrow[draw=black]{>}}}},
    fermionbar/.style={draw=black, postaction={decorate},
        decoration={markings,mark=at position .55 with {\arrow[draw=black]{<}}}},
    fermionnoarrow/.style={draw=black},
    gluon/.style={decorate, draw=black,
        decoration={coil,amplitude=4pt, segment length=5pt}},
    scalar/.style={dashed,draw=black, postaction={decorate},
        decoration={markings,mark=at position .55 with {\arrow[draw=black]{>}}}},
    scalarbar/.style={dashed,draw=black, postaction={decorate},
        decoration={markings,mark=at position .55 with {\arrow[draw=black]{<}}}},
    scalarnoarrow/.style={dashed,draw=black},
    electron/.style={draw=black, postaction={decorate},
        decoration={markings,mark=at position .55 with {\arrow[draw=black]{>}}}},
	bigvector/.style={decorate, decoration={snake,amplitude=4pt}, draw},
}
\tikzstyle{block} = [draw, rectangle,
    minimum height=3em, minimum width=6em]

\definecolor{greeen}{rgb}{0.03,0.84,0.13}
\definecolor{test}{rgb}{0.03,0.74,0.33}
\definecolor{viol}{rgb}{0.44,0,0.94}
\definecolor{or}{rgb}{0.95,0.65,0}

\definecolor{MyDarkBlue}{rgb}{0.1, 0.1, 0.8} 
\definecolor{MyLightBlue}{rgb}{0.22,0.51,0.9}

\begin{document}

\baselineskip=18pt

\numberwithin{equation}{section}

\vspace*{-0.2in}
\begin{flushright}
OSU-HEP-18-06\\
MI-TH-181\\
{UMD-PP-018-08}
\end{flushright}
\vspace{0.5cm}
\begin{center}

\thispagestyle{empty}

{\Large{\bf{A Theory of \boldmath{${R(D^*,D)}$} Anomaly\\[0.1in] With Right-Handed Currents}}}
\vspace{1cm}

\renewcommand{\thefootnote}{\fnsymbol{footnote}}
\centerline{
{}~{\bf K.S. Babu}$^{a,}$\footnote{E-mail: \textcolor{MyLightBlue}{babu@okstate.edu}},{}~
{\bf Bhaskar Dutta}$^{b,}$\footnote{E-mail: \textcolor{MyLightBlue}{dutta@physics.tamu.edu}} and
{}~{\bf Rabindra N. Mohapatra}$^{c,}$\footnote{E-mail: \textcolor{MyLightBlue}{rmohapat@physics.umd.edu}}
}

\vspace{0.5cm}

$^{a}${\it\small Department of Physics, Oklahoma State University,}\\ {\it\small Stillwater, OK, 74078, USA}
\vspace*{0.1in}

$^{b}${\it\small Mitchell Institute for Fundamental Physics and Astronomy,}\\ {\it\small
Department of Physics \& Astronomy, Texas A \& M University, College Station, TX, 77845, USA}
\vspace*{0.1in}

$^{c}${\it\small Maryland Center for Fundamental Physics, Department of Physics,}\\
{\it\small University of Maryland, College Park, MD 20742, USA}
\end{center}

\renewcommand{\thefootnote}{\arabic{footnote}}

\bigskip

\begin{abstract}
{\footnotesize
We present an ultraviolet complete theory for the $R(D^*)$ and $R(D)$ anomaly in terms of a low mass $W_R^\pm$ gauge boson of a class of left-right symmetric models. These models, which are based on the gauge symmetry $SU(3)_c \times SU(2)_L \times SU(2)_R \times U(1)_{B-L}$, utilize vector-like fermions to generate quark and lepton masses via a universal seesaw mechanism. A parity symmetric version as well as an asymmetric version are studied. A light sterile neutrino emerges naturally in this setup, which allows for new decay modes of $B$-meson via right-handed currents.  We show that these models can explain $R(D^*)$ and $R(D)$ anomaly while being consistent with LHC and LEP data as well as low energy flavor constraints arising from $K_L-K_S, B_{d,s}-\bar{B}_{d,s}$, $D-\bar{D}$ mixing, etc., but only for a limited range of the $W_R$ mass: $1.2\, (1.8)~{\rm TeV} \leq M_{W_R}\leq 3~ {\rm TeV}$ for parity asymmetric (symmetric) Yukawa sectors. The light sterile neutrinos predicted by the model may be relevant for explaining the MiniBoone and LSND neutrino oscillation results. The parity symmetric version of the model provides a simple solution to the strong CP problem without relying on the axion. It also predicts an isospin singlet top partner with a mass $M_T = (1.5-2.5)$ TeV.
}
 \end{abstract}

\clearpage

\section{Introduction}

The observations by  BaBar \cite{babar1,babar2}, Belle~\cite{belle1,belle2,belle3} and LHCb~\cite{lhcb} experiments of  deviations in the ratio of $B$ meson decays $R(D^*) = \frac{\Gamma(B\to D^*\tau\nu)}{\Gamma(B\to D^*\ell\nu)}$ and
$R(D) = \frac{\Gamma(B\to D\tau\nu)}{\Gamma(B\to D\ell\nu)}$ from their standard model predictions at $\sim 4\sigma$ level have posed quite a theoretical challenge. Recently LHCb has released its first measurement of the ratio of branching ratios $\frac{{\cal B}(B^+_c\to J/\psi \tau \nu)}{{\cal B}(B^+_c\to J/\psi \mu \nu)}$~\cite{Jpsi} which also differs from its standard model prediction at the 2 $\sigma$ level, apparently supporting the above anomaly.  An intriguing possibility discussed recently~\cite{He1,shakya,buckley} is that there may be additional contributions only to the $D \tau \nu$ decay mode of $B$-meson mediated by a low mass $SU(2)_L$ singlet $W'$ boson which exclusively couples to $\bar{b}_R\gamma_\mu c_R$ and $\bar{\tau}_R\gamma_\mu\nu_R$ currents with a gauge coupling $g_R$. For $g_R$ equal to the weak $SU(2)_L$ gauge coupling $g_L$, and with no mixing angle suppression in the $\bar{b}_R\gamma_\mu c_R$ vertex, resolving the $R(D, D^*)$ anomaly would require that $M_{W'}\simeq 700$ GeV. The question then is what kind of an ultraviolet complete theory would lead to such an interaction.  It would be of great interest to see if such a $W'$ can be identified with a low mass right handed $W_R^\pm$ boson of left-right symmetric theories~\cite{LR} discussed extensively in the literature. We explore this question in this paper.

A major hurdle that any left-right symmetric embedding of low mass $W'$ should overcome is the current lower bound on $M_{W_R}$  from flavor changing neutral current observables such as $K_L-K_S$, $B_s-\bar{B}_s$ and $B_d-\bar{B}_d$ mixings~\cite{FCNC,Zhang}, as well as the direct $W_R$ search limits at the LHC \cite{miha}. Furthermore, since in simple left-right models there is a relation between the masses of $W_R$ and $Z_R$, e.g. $M_{Z_R} \simeq 1.7 \, (1.2) M_{W_R}$ in parity symmetric models with Higgs triplet (doublet) used for $SU(2)_R$ breaking, one has to reconcile a low mass $Z_R$ with current limits from LEP and LHC searches \cite{fox}. Indeed, consistency with these constraints would prevent an explanation of the anomaly in terms of $W_R$ from the standard formulation of left-right symmetric models with parity \cite{LR,FCNC,Zhang,miha} or without parity \cite{Langacker}. We will have more to say on this conclusion later (see discussions in Sec. 4.1).

 We focus here on a variant formulation of left-right (LR) symmetric models which introduces vector-like fermions for quark and lepton mass generation and show that such a setup can overcome the hurdles mentioned above. These models, which are based on the standard LR gauge group $SU(3)_c\times SU(2)_L\times SU(2)_R\times U(1)_{B-L}$, have a very simple Higgs sector -- one $SU(2)_L$ doublet $\chi_L$ and one $SU(2)_R$ doublet $\chi_R$.  Vector-like fermions $(U_a,\, D_a,\, E_a,\, N_a)$ with $a=1-3$ transforming as singlets of $SU(2)_L$ and $SU(2)_R$ and with electric charges $(2/3,\, -1/3,\, -1,\, 0)$ are needed to generate fermion masses, which arise via a  ``universal seesaw" mechanism \cite{berez,chang,wali,bm}.  These vector-like fermions can have gauge invariant bare masses, with $N_{La}$ and $N_{Ra}$ having both Dirac and Majorana masses.

Although not very minimal in the fermionic content, these models do provide certain advantages.  First, the Higgs sector is very minimal, with the physical spectrum consisting only of two neutral scalars, one of which being the 125 GeV Standard Model--like Higgs boson.  Second, owing to the quadratic dependence of the light fermion masses on the Yukawa couplings $Y_i$, the values of $Y_i$ needed to explain the hierarchy in fermion masses can be in the range $Y_i = (10^{-3} - 1)$ as opposed to $Y_i = (10^{-6} - 1)$ in the standard left-right symmetric model (or in the standard model). This follows as the fermion masses in these models are given as $m_i \sim Y_i^2 \kappa_L \kappa_R/M_i$, assuming parity, where $\kappa_{L,R}$ are the vacuum expectation values (VEVs) of the $SU(2)_{L,R}$ doublet fields $\chi_{L,R}$. Third, these models provide naturally light sterile neutrinos, which leads to the possibility of right-handed currents in meson decays, and which may play a role in understanding the MiniBoone \cite{MiniBoone} and LSND \cite{LSND} neutrino oscillation results.\footnote{Models with light sterile neutrinos introduced to explain the MiniBoone and LSND anomalies would appear to be in conflict with the number of effective neutrino species inferred from $\Lambda$CDM cosmology, especially from Planck data \cite{Planck}. A possible way around this within $\Lambda$CDM is to postulate secret self-interactions of these sterile neutrinos \cite{br}.}
And fourth, these models provide a simple solution to the strong CP problem based on parity symmetry alone, without the need for an axion.  The QCD $\theta$ parameter is zero, and the determinant of the tree-level quark mass matrix is real, both owing to parity symmetry \cite{bm}.  Small and calculable $\overline{\theta}$ is induced in the model only at the two-loop level, which is consistent with neutron electric dipole moment constraints \cite{bm,hall}.

While we do emphasize the parity symmetric version of the universal seesaw models, in addressing the $R(D^*,D)$ anomaly, we shall also deviate from the requirement of exact parity.  Some of the motivations, but not all, quoted above will not be valid in this parity asymmetric scenario.
In this case we use a partial quark and lepton seesaw, where the seesaw is effective only for a subset of quark and charged lepton families \cite{BEM}.  This enables us to straightforwardly evade the most stringent flavor constraints and still be able to explain $R(D^*,D)$ results. The same result is also achieved in a parity symmetric scenario where parity is broken softly and spontaneously
without relying on partial quark and lepton seesaw, but with a different flavor structure in the Yukawa sector.

The main results of the paper are the following.  (i) A low mass $W_R$ is needed to explain $R(D^*,D)$ anomaly consistent with LHC and LEP constraints, with the mass range given by  $1.2 \,(1.8)~{\rm TeV} \leq M_{W_R}\leq 3~ {\rm TeV}$ in the parity asymmetric (symmetric) version. (ii) The widths of the $W_R$ and $Z_R$ turn out to be relatively large, $\Gamma(W_R,Z_R)/M_{W_R,Z_R} \geq 20\%$, when $R(D^*,D)$ anomaly is explained, which helps us reconcile their low masses with LHC searches.
(iii) Explaining $R(D^*, D)$ observations imposes stringent constraints on the flavor structure of the model in the right-handed sector. (iv) In the parity symmetric version, the strong CP problem is solved without the need for an axion. This model predicts a vector-like top partner quark to have a mass $M_T = (1.5-2.5)$ TeV. In the parity asymmetric case, the flavor structure we adopt leads to a limit $M_i<2.5$ TeV for several vector-like quarks and leptons.

Our model setup differs significantly from that of Ref. \cite{He1} which also uses right-handed currents in that all three families transform under $SU(2)_R$ in our case, as opposed to only the third family in Ref. \cite{He1}.  The models of Ref. \cite{shakya} have new vector-like fermions (and not the SM fermions) transforming under $SU(2)_R$.  The model of Ref. \cite{buckley} also assumes vector-like fermions transforming under $SU(2)_R$, with the SM fermions acquiring $SU(2)_R$ charge only via mixing with these vector-fermions.  The universal seesaw setup that we pursue here is independently motivated, as noted earlier, especially for the solution it provides for the strong CP problem based on parity symmetry. There are of course other popular explanations for the $R(D^*,D)$ anomaly, in terms of  leptoquarks \cite{lepto},   a $W'$ that couples to left-handed fermion fields \cite{W',Abdullah:2018ets}, supersymmetry~\cite{susy} and extra dimension~\cite{extradimension}.  Explanations in terms of additional scalars \cite{2H} appear to be in tension with the branching ratio constraint ${\cal B}(B_c \rightarrow \tau \nu$) \cite{grinstein,akeroyd}, but the significance of the anomaly may still be reduced.

This paper is organized as follows. In Sec. 2 we describe the details of the universal seesaw version of the LR model. In Sec. 3 we develop a parity asymmetric version of the model and identify a suitable flavor structure for $R(D^*,D)$ anomaly.  Here we also discuss
how the constraints on the model from flavor changing observations such as $K_L-K_S$ mass difference, $D-\bar{D}$, $B_{d,s}-\bar{B}_{d,s}$ transitions, electroweak precision data, etc. are satisfied. In Sec. 4, we develop a parity symmetric versions, which also solves the strong CP problem, which we briefly review.  In Sec. 5 we show how the model explains the $R(D^*,D)$ anomaly. In Sec. 6 we discuss how low mass $W_R$ and $Z_R$ required for explaining $R(D^*,D)$ evades the LHC and LEP constraints.  We comment on some cosmological and astrophysical constraints on the model in Sec. 7.  Finally, in  Sec. 8 we offer some theoretical comments on the model and conclude.

\section{ Left-right symmetric models with universal seesaw}

 We focus on a class of left-right symmetric models based on the gauge group $SU(3)_c\times SU(2)_L\times SU(2)_R\times U(1)_{B-L}$ where fermion masses are induced through a universal seesaw mechanism \cite{berez,chang,wali,bm}. This setup enables one to define Parity ($P$) as a spontaneously broken symmetry. Imposing $P$ would strongly constrain the gauge and Yukawa couplings of the left-handed and right-handed fermions. We consider two versions of the model: One  without parity where the couplings in the left-handed and right-handed fermion sectors are arbitrary and unrelated to each other; and a second one where parity is a softly broken symmetry where the left-handed and right-handed Yukawa couplings are identified.  We shall see that both versions can explain the $R(D^*,D)$ anomaly, but with different choices of flavor structure. These models have the usual standard model fermions plus the right-handed neutrinos needed to complete the right-handed lepton doublet. In contrast with the usual left-right models, the universal seesaw version has four extra sets of vector-like fermions which are $SU(2)_L\times SU(2)_R$ singlets, denoted as $(U_a, D_a, E_a, N_a)$. The chiral fermions are assigned to the gauge group as follows ($i=1-3$ is the family index):
\begin{eqnarray}
Q_{L,i}\left({  3}, { 2}, { 1}, +\frac{1}{3}\right) &=&  \left(\begin{array}{c}u_L\\d_L \end{array}\right)_i ,  ~~~~~
Q_{R,i}\left({ 3}, { 1}, { 2}, +\frac{1}{3}\right)  =  \left(\begin{array}{c}u_R\\d_R \end{array}\right)_i ,
\nonumber \\
\psi_{L,i}\left({  1}, {2}, { 1}, -1 \right)  &=&   \left(\begin{array}{c}\nu_L \\ e_L \end{array}\right)_i , ~~~~~
\psi_{R,i}\left({ 1}, { 1}, { 2}, -1 \right)  =  \left(\begin{array}{c} \nu_R \\ e_R \end{array}\right)_i ~.
\label{lrSM}
\end{eqnarray}
The three families ($a=1-3$) of vector-like fermions have the following gauge quantum numbers for both left-handed and right-handed chiralities:
\begin{eqnarray}
 U_{a} (3, 1,1,+\frac{4}{3}),~~~~ D_{a} (3,1,-\frac{2}{3}),~~~~ E_{a} (1,1,1,-2),~~~~N_a (1,1,1,0)~.
 \end{eqnarray}

  The Higgs sector is very simple consisting of a left-handed and a right-handed doublet:
  \begin{eqnarray}
  \chi_L(1,2,1,+1) = \left(\begin{matrix}\chi_L^+ \\ \chi_L^0   \end{matrix}  \right),~~~~  \chi_R(1,1,2,+1) = \left(\begin{matrix}\chi_R^+ \\ \chi_R^0   \end{matrix}  \right)~.
    \end{eqnarray}
  Note in particular that there are no bidoublet scalar fields in the model. The physical Higgs boson spectrum has just two neutral scalars,
  $\sigma_L = {\rm Re}(\chi_L^0)/\sqrt{2}$ and $\sigma_R = {\rm Re}(\chi_R^0)/\sqrt{2}$ which mix, with the SM-like Higgs boson of mass 125 GeV identified as primarily $\sigma_L$.  The charged $\chi_{L,R}^\pm$ and neutral pseudo-scalar bosons Im$(\chi_{L,R}^0)/\sqrt{2}$ are eaten up by the $(W_L^\pm, W_R^\pm)$ and the $(Z_L^0, Z_R^0)$ gauge bosons. We shall denote the vacuum expectation values of the neutral members of $\chi_{L,R}$ as
  \begin{equation}
 \left\langle\chi^0_{L}\right\rangle=\kappa_{L};~~~ \left\langle\chi^0_{R}\right\rangle=\kappa_{R}~
 \end{equation}
with $\kappa_L \simeq 174$ GeV.

Among the charged gauge bosons, $W_L^\pm$ and $W_R^\pm$ do not mix at tree-level. Their masses are given by
\begin{equation}
 M^2_{W^\pm_L}~=~\frac{g^2_L\kappa^2_L }{2},~~~ M^2_{W^\pm_R}~=~\frac{g^2_R\kappa^2_R }{2}~.
\end{equation}
In the neutral gauge boson sector, the states $(W_{3L},\,W_{3R},\,B)$ will mix (where $B$ denotes the $B-L$ gauge boson).  The photon filed $A_\mu$ remains massless, while the two orthogonal fields $Z_L$ and $Z_R$ mix.  The compositions of these fields, in a certain convenient basis, take the form:
\begin{eqnarray}
A^\mu &=& \frac{g_L g_R B^\mu + g_B g_R W_{3L}^\mu + g_L g_B W_{3R}^\mu}{\sqrt{g_B^2(g_L^2+g_R^2) + g_L^2 g_R^2}} \nonumber \\
Z_R^\mu &=& \frac{g_B B^\mu - g_R W_{3R}^\mu}{\sqrt{g_R^2+g_B^2}} \nonumber \\
Z_L^\mu &=& \frac{g_B g_R B^\mu-g_L g_R\left(1+\frac{g_B^2}{g_R^2}\right) W_{3L}^\mu + g_B^2 W_{3R}^\mu}{\sqrt{g_B^2+g_R^2} \sqrt{g_B^2+g_L^2+\frac{g_B^2 g_L^2}{g_R^2}}}~
\end{eqnarray}
with the $Z_L-Z_R$ mixing matrix given as
\begin{eqnarray}
{\cal M}^2_{Z_L-Z_R} =
\frac{1}{2} \, \left( \begin{matrix} (g_Y^2+g_L^2)\, \kappa_L^2 & g_Y^2 \sqrt{\frac{g_Y^2 + g_L^2}{g_R^2-g_Y^2}} \,\kappa_L^2 \\
 g_Y^2 \sqrt{\frac{g_Y^2 + g_L^2}{g_R^2-g_Y^2}}\, \kappa_L^2 & \frac{g_R^4}{g_R^2-g_Y^2}\, \kappa_R^2 + \frac{g_Y^4}{g_R^2-g_Y^2}\, \kappa_L^2      \end{matrix} \right)~.
 \label{ZLR}
\end{eqnarray}
Here $g_R$ and $g_{B}$ are the $SU(2)_R$ and $U(1)_{B-L}$ gauge couplings which are
 related to the hypercharge coupling $g_Y$ through the formula that embeds $Y$ within $SU(2)_R \times U(1)_{B-L}$:
\begin{equation}
\frac{Y}{2} = T_{3R} + \frac{B-L}{2} ~~~~~ \Rightarrow~~~~~ g_Y^{-2} = g_R^{-2} +g_B^{-2}~.
\label{embed}
\end{equation}
We have eliminated $g_B$ in favor of $g_Y$ in Eq. (\ref{ZLR}).

The physical states and their masses are given by
\begin{eqnarray}
Z_1 &=& \cos \xi\, Z_L - \sin\xi\, Z_R, ~~~~Z_2 = \sin\xi\, Z_L + \cos\xi \,Z_R, \nonumber \\
M_{Z_1}^2 &\simeq& \frac{1}{2} (g_Y^2+g_L^2)\, \kappa_L^2,~~~~~~~~M_{Z_2}^2 \simeq
\frac{g_R^4}{g_R^2-g_Y^2}\, \kappa_R^2 + \frac{g_Y^4}{g_R^2-g_Y^2}\, \kappa_L^2
\end{eqnarray}
with the mixing angle $\xi$ given approximately by
\begin{equation}
\xi \simeq \frac{g_Y^2}{g_R^4}\, \sqrt{(g_L^2+g_Y^2)(g_R^2-g_Y^2)} \, \frac{\kappa_L^2}{\kappa_R^2}~.
\end{equation}
Here $Z_1$ is identified as the $Z$ boson.
As it turns out, $\xi$ is very small for typical parameters that would be used to explain $R(D^*,D)$ anomaly.  A benchmark point is $M_{W^\pm_R} = 2$ TeV, and $g_R = 2$. This corresponds to $\kappa_R = 1.4$ TeV, in which case $\xi \simeq 1.8 \times 10^{-4}$.  If we choose instead, $g_R = g_L \simeq 0.65$, $\xi \simeq 4.7 \times 10^{-4}$, again for $M_{W^\pm_R} = 2$ TeV. Such a small value of $\xi$ has very little impact in our analysis.  For example, the new contribution to the electroweak parameter $\alpha T$ is given by $\alpha T \simeq \xi^2 M_{Z_R}^2/M^2_{Z_L} \simeq 1.6 \times 10^{-5}$ (for $g_R = 2$), well below the experimental limits.  The decays of $Z_2$ into diboson channels, viz., $Z_2 \rightarrow W_L^+ W_L^-$ and $Z_2 \rightarrow Z_1 h$ (where $h$ is the 125 GeV Higgs boson) will proceed through the $Z_L-Z_R$ mixing, however, with non-negligible partial widths. We shall take $Z_L-Z_R$ mixing into account in discussing such diboson decays in Sec. 6.

The Higgs potential of the model is given by
\begin{eqnarray}
V = -(\mu_L^2 \chi_L^\dagger \chi_L +\mu_R^2 \chi_R^\dagger \chi_R) +  \frac{\lambda_{1L}}{2}(\chi_L^\dagger \chi_L)^2+ \frac{\lambda_{1R}}{2}(\chi_R^\dagger \chi_R)^2   + \lambda_2 (\chi_L^\dagger \chi_L)(\chi_R^\dagger \chi_R)~.
\label{V}
\end{eqnarray}
If Parity symmetry is assumed, we would have $\lambda_{1L} = \lambda_{1R} \equiv \lambda_1$.  We shall allow for soft breaking of $P$, in which case the quadratic terms $\mu_L^2 \neq \mu_R^2$.\footnote{In Ref. \cite{hall} it has been shown that soft breaking of $P$ in the Higgs potential is not necessary if $\kappa_R \sim 10^{11}$ GeV. For explaining $R(D^*,D)$ via $W_R$, our setup requires $\kappa_R \sim 2$ TeV, in which case soft breaking of $P$ would be needed.}
The physical Higgs spectrum is obtained from the $\sigma_L-\sigma_R$ mixing matrix ($\sigma_L = {\rm Re}(\chi_L^0)/\sqrt{2}$, $\sigma_R = {\rm Re}(\chi_R^0)/\sqrt{2}$) given by
\begin{eqnarray}
{\cal M}^2_{\sigma_{L,R}} = \left[\begin{matrix}2 \lambda_{1L} \kappa_L^2 & 2 \lambda_2 \kappa_L \kappa_R \\ 2 \lambda_2 \kappa_L \kappa_R & 2 \lambda_{1R} \kappa_R^2   \end{matrix}\right]~.
\end{eqnarray}
The eigenstates and the respective mass eigenvalues are given by
\begin{eqnarray}
h &=& \cos \zeta \, \sigma_L - \sin\zeta \, \sigma_R,~~~~~ H = \sin\zeta \, \sigma_L + \cos\zeta \, \sigma_R, \nonumber \\
M_h^2 &\simeq & 2 \lambda_{1L}\left(1- \frac{\lambda_2^2}{\lambda_{1L}\lambda_{1R}}  \right) \kappa_L^2,~~~~~~M_H^2 \simeq 2 \lambda_{1R} \kappa_R^2
\end{eqnarray}
with the mixing angle $\zeta$ given by
\begin{equation}
\tan 2 \zeta = \frac{2 \lambda_2 \kappa_L \kappa_R}{(\lambda_{1R}\kappa_R^2-\lambda_{1L} \kappa_L^2)}~.
\end{equation}
We note that boundedness of the potential requires
\begin{equation}
\lambda_{1L} \geq 0, ~~~\lambda_{1R} \geq 0,~~~\lambda_2 \geq -\sqrt{\lambda_{1L}\lambda_{1R}}~.
\end{equation}
The mixing angle $\zeta$ will be relevant for the decays $Z_2 \rightarrow Z_1 +h$, $Z_2 \rightarrow Z_1 + H$, and $Z_2 \rightarrow h + H$, the latter two when kinematically allowed.

Turning to the fermion masses, the Yukawa couplings and the mass terms in the charged sector have the form
\begin{eqnarray}
{\cal L}_{\rm Yuk} &=& Y_U \overline{Q}_L \tilde{\chi}_L U_R + Y_U' \overline{Q}_R \tilde{\chi}_R U_L + M_U \overline{U}_L U_R \nonumber \\
&+& Y_D\overline{Q}_L \chi_L D_R + Y_D' \overline{Q}_R \chi_R D_L + M_D \overline{D}_L D_R \nonumber \\
&+& Y_E \overline{\psi}_L \chi_L E_R + Y_E'\overline{\psi}_R \chi_R E_L + M_E \overline{E}_L E_R + h.c.
\label{Yuk}
\end{eqnarray}
Here $\tilde{\chi}_{L,R} = i\tau_2 \chi_{L,R}^*$.  When Parity symmetry is imposed, under $P$
the fermion and scalar fields transform as follows:
\begin{eqnarray}
Q_L \leftrightarrow Q_R,~~\psi_L \leftrightarrow \psi_R,~~~U_L \leftrightarrow U_R,~~~D_L \leftrightarrow D_R,~~~E_L \leftrightarrow E_R,~~~
\chi_L \leftrightarrow \chi_R~.
\end{eqnarray}
Simultaneously, $W_L \leftrightarrow W_R$.  The parameters in Eq. (\ref{Yuk}) would then satisfy the following conditions:
\begin{eqnarray}
Y_U = Y'_U,~~~Y_D = Y'_D,~~~Y_E = Y'_E,~~~M_U = M_U^\dagger,~~~M_D = M_D^\dagger,~~~M_E = M_E^\dagger
\end{eqnarray}
along with $g_L = g_R$ on the $SU(2)_L$ and $SU(2)_R$ gauge couplings.  If $P$ is a softly broken symmetry, then the hermiticity conditions on $M_{U,D,E} = M_{U,D,E}^\dagger$ are not required.  In the $P$ symmetric models that we discuss, we shall take $M_{U,D,E} \neq M_{U,D,E}^\dagger$.

The $6 \times 6$ mass matrices in the up-quark, down-quark and charged lepton sectors arising from Eq. (\ref{Yuk}) take the form
\begin{eqnarray}
{\cal M}_{U,D,E}~=~\left(\begin{array}{cc}0 & Y_{U,D,E}\kappa_L\\Y^{\prime \dagger}_{U,D,E} \kappa_R & M_{U,D,E}\end{array}\right)~.
\label{mass}
\end{eqnarray}
Here the basis is $(u,c,t,U,C,T)$ in the up-quark sector, $(d,s,b,D,S,B)$ in the down-quark sector and $(e,\mu,\tau,E_1,E_2,E_3)$ in the charged lepton sector, with the left-handed fields multiplying the matrix from the left and the right-handed fields multiplying from the right in Eq. (\ref{mass}).  (We use capitalized $(U,C,T)$ for the heavy vector-like up-quarks and so forth.) If the determinants of the bare mass terms $M_{U,D,E}$ in Eq. (\ref{mass}) are all nonzero, the light eigenvalues will be given by $m_i \sim Y_i Y_i'\kappa_L \kappa_R/M_i$ (ignoring generation mixing).  This is the universal seesaw mechanism.

If parity is imposed as a softly broken symmetry, then $Y'_{U,D,E} = Y_{U,D,E}$ in Eq. (\ref{mass}).  Note that $M_{U,D,E}$ can  be non-hermitian as these terms break $P$ only softly. The QCD parameter $\theta $ can be set to zero by virtue of Parity. Furthermore, the VEVs $\kappa_{L,R}$ can be made real by $SU(2)_{L,R}$ gauge rotations. This is a crucial point possible only because the Higgs sector is very simple.  With parity, then, we have the determinants of ${\cal M_U}$ and ${\cal M_D}$ being real.  This leads to the result $\overline{\theta} = 0$ at tree-level \cite{bm}.  Even with $M_{U,D} \neq M_{U,D}^\dagger$, nonzero $\overline{\theta}$ is induced via two-loop diagrams, which turn out to be of order $10^{-10}$ \cite{bm,hall}. Thus the parity symmetric version of the model provides a simple solution to the strong CP problem without the need for an axion.

The soft breaking of $P$ in the scalar mass terms of Eqs. (\ref{V}) and in the fermion mass terms of (\ref{Yuk}) can be understood as a spontaneous breaking at a higher scale.  A Parity odd real singlet scalar field $S$ can couple to the Higgs fields and the fermion fields~\cite{CMP}. Under $P$, $S \rightarrow - S$. These couplings, along with the $P$ symmetric bare couplings are given by:
\begin{eqnarray}
{\cal L}_S &=& \mu_0^2(\chi_L^\dagger \chi_L + \chi_R^\dagger \chi_R)  + \left \{M_U^0 \overline{U}_L U_R + M_D^0 \overline{D}_L D_R + M_E^0 \overline{E}_L E_R + h.c. \right\} + \nonumber \\
&+& \mu_1 S (\chi_L^\dagger \chi_L - \chi_R^\dagger \chi_R)  +
 S \left\{Y_U^S \overline{U_L} U_R + Y_D^S \overline{D}_L D_R + Y_E^S \overline{E}_L E_R + h.c. \right \}
\end{eqnarray}
with $M_{U,D,E}^0 = M_{U,D,E}^{0 \dagger}$ and $Y_{U,D,E}^S = - Y_{U,D,S}^{S \dagger}$.  Once $S$ acquires a vacuum expectation value, the mass parameters of Eq. (\ref{V}) will be generated with  $\mu_L^2 = \mu_0^2 + \mu_1 \left\langle S \right\rangle$ and $\mu_R^2 = \mu_0^2 - \mu_1 \left\langle S \right\rangle$. Similarly, in Eq. (\ref{Yuk}) non-hermitian mass matrices will be generated given by $M_{U,D,E} = M_{U,D,E}^0 + Y_{U,D,SE}^S \left\langle S \right\rangle$. Although we shall not explicitly make use of the parity odd singlet scalar $S$, this argument shows the consistency of treating $P$ as a softly broken symmetry.

As for the neutrinos, the Yukawa Lagrangian is given by
\begin{eqnarray}
{\cal L}_{\rm Yuk}^\nu &=& Y_\nu \overline{\psi}_L \tilde{\chi}_L N_R + Y_\nu' \overline{\psi}_R \tilde{\chi}_R N_L + \tilde{Y}_\nu \overline{\psi}_L \tilde{\chi}_L N^c_R + \tilde{Y}'_\nu \overline{\psi}_R \tilde{\chi}_R N^c_L \nonumber \\
&+& M_N \overline{N}_L N_R + \mu_L N_L^T C N_L + \mu_R N_R^T C N_R + h.c.
\end{eqnarray}
Note the presence of a Dirac mass term $M_N$ and Majorana mass terms $\mu_L$ and $\mu_R$.  The $12 \times 12$ Majorana mass matrix in the basis $(\nu_i, \nu^c_i, N_i, N^c_i)$ -- with all fields taken to be left-handed and the matrices to be taken real for simplicity-- is given by
\begin{eqnarray}
{\cal M}_\nu = \left(\begin{matrix}0 & 0 & Y_\nu \kappa_L & \tilde{Y}_\nu \kappa_L \\ 0 & 0 & Y_\nu' \kappa_R & \tilde{Y}_\nu' \kappa_R \\
Y_\nu^T \kappa_L & Y_\nu'^T \kappa_R & \mu_L & M_N \\ \tilde{Y}_\nu^T \kappa_L & \tilde{Y}_\nu'^T \kappa_R & M_N^T & \mu_R
  \end{matrix}\right)~.
  \label{nu}
\end{eqnarray}
Under parity, $N_L \leftrightarrow N_R$, which would imply $Y_\nu = Y_\nu'$, $\tilde{Y}_\nu = \tilde{Y}_\nu'$, $\mu_L = \mu_R$ and $M_N = M_N^\dagger$. The last two relations will not hold when $P$ is softly broken, and therefore will not be assumed.  An interesting feature of this mass matrix is that it naturally leads to light sterile neutrinos.  Thus this setup allows for the kinematic decay of $B$ meson into these sterile neutrinos.  To see the emergence of light sterile neutrinos, consider decoupled generations and focus on one such generation.  With $M_N \sim \mu_{L,R}$, two eigenvalues of the mass matrix in Eq. (\ref{nu}) will be of order $M_N$, while the $\nu^c$ state will have a mass of order $Y_\nu^2 \kappa_R^2/M_N$. The lighter $\nu$ state has a mass of order $Y_\nu^2 \kappa_L^2/M_N$.  In order to explain the smallness of the light neutrino masses, $M_N \gg \kappa_R$ would be preferred. We see that with $\kappa_R \ll M_N$, the mass of $\nu_R$ is much smaller than $\kappa_R$, and can be in the sub-MeV range.  It is also possible that $\mu_L \sim \mu_R \gg M_N$, with additional symmetries keeping the bare Dirac mass $M_N$ of order TeV, just as the charged fermion bare mass terms. The Majorana mass terms $\mu_{L,R}$, which could obey different selection rule, need not be protected by such symmetries and can be of order $10^{10}$ GeV or so.  Again, $\nu_R$ mass will be much smaller than $\kappa_R$ and may be in the sub-MeV or even in the eV range.

It is intriguing to note that the $\nu_L$ to $\nu_R$ mass ratio is approximately given as $\kappa_L^2/\kappa_R^2$, provided that there is no special flavor structure in $M_N$, $\mu_L$ and $\mu_R$.  From a fit to the $R(D^*,D)$ anomaly we shall find the ratio $(\kappa_L/\kappa_R)^2 \sim 1/60$, in which case the sterile neutrino mass comes out to be near 3 eV, if we use the active neutrino mass to be 0.05 eV from atmospheric neutrino oscillation data, assuming normal mass ordering.  This is in the right range for explaining the MiniBoone and LSND neutrino oscillation data. (See however, the cosmological caveat noted in footnote 4.) We shall allow for this possibility, as well as the case where the $\nu_R$ states are heavier (except for $\nu_{\tau_R}$) as a result of possible structures in the bare mass terms in Eq. (\ref{nu}).

It should be noted that there is an option to remove the singlet fields $N_L$ and $N_R$ from the theory, in which case the neutrino will be pure Dirac particle \cite{bh}.  It is also possible to provide the light active neutrinos small Majorana masses by introducing a $\Delta_L(1,3,1,+2)$ field via type-II seesaw mechanism. A parity partner $\Delta_R(1,1,3,+2)$ can acquire a small induced VEV and generate small Majorana masses for the $\nu_R$ fields \cite{BEM}.  For concreteness we shall adopt the mass matrix of Eq. (\ref{nu}) for neutrino mass generation, and not these variant schemes.

\section{ Parity asymmetric flavor structure without FCNC}

In this section we develop a scenario without assuming parity symmetry that  explains the $R(D^*,D)$ anomaly consistent with other flavor violation constraints. When parity symmetry is not assumed, the left-handed and right-handed fermions can have independent Yukawa couplings.  Thus this version of the model has more freedom, compared to the parity symmetric version that will be developed in the next section. We choose in this case a specific flavor structure motivated on the one hand by $R(D^*,D)$ anomaly and by the need to eliminate large flavor violation that could arise in this setup on the other hand.

Suppose that parity is not a good symmetry.  Then the seesaw mechanism may be only effective partially, which happens when ${\rm Det} (M_{U,D,E})=0$.  In this case, the seesaw formula breaks down for some fermions. To see this in detail, let us work in a basis where the fermion mass matrices are block-diagonal and $M_{U,D,E}$ are diagonal. If any one of the diagonal elements of $M_{U,D,E}$ is zero we have ${\rm Det} (M_{U,D,E})=0$ . In that case, the fermion fields split into two groups: for a generation for which the vector-like bare mass term vanishes, there is a heavy fermion with mass $\sim Y^\prime\kappa_R$ which is coupled to $W_R$, and a light fermion with masses $\sim Y \kappa_L$ coupling only to $W_L$. For the generations for which $M_F\neq 0$, there is a light fermion whose mass is given by the seesaw formula $m_i \sim Y_i Y_i'\kappa_L\kappa_R/M_i$ and  which couples to both $W_L$ and $W_R$. It is this property of partial seesaw which helps us to have a $W_R$ couple exclusively to $\overline{b}_R \gamma_\mu c_R$ and $\overline{\tau}_R \gamma_\mu \nu_R$ in the parity asymmetric case.

As an explicit example, we make the following choice. In the quark sector, for the various blocks of the mass matrices of Eq. (\ref{mass}) we choose:
\begin{eqnarray}
Y_U &=&  V_L^\dagger Y_U^{\rm diag},~~~Y_U'= V_R^\dagger Y_U^{\prime \,{\rm diag}}, ~~M_U = {\rm diag}(0, M_2,0)~\nonumber\\
Y_D &=& Y_D^{\rm diag},~~~ Y_D' = Y_D^{\prime\, {\rm diag}},~~~ M_D={\rm diag}(0,0,M_3)
\label{flip}
\end{eqnarray}
Here $Y_U^{\rm diag} = {\rm diag}(Y_1^u, Y_2^u, Y_3^u)$, $Y_U^{\prime \,{\rm diag}} = {\rm diag}(Y_1^{\prime u}, Y_2^{\prime u}, Y_3^{\prime u})$,
$Y_D^{\rm diag} = {\rm diag}(Y_1^d, Y_2^d, Y_3^d)$ and $Y_D^{\prime\, {\rm diag}} = {\rm diag}(Y_1^{\prime d}, Y_2^{\prime d}, Y_3^{\prime d})$ are arbitrary diagonal matrices. $V_L$ is the left-handed CKM matrix, while $V_R$ is the right-handed CKM matrix, which is unrelated to $V_L$.  $V_L$ is chosen to fit the CKM matrix elements, while we choose $V_R$ to have the form:
\begin{eqnarray}~V_R=\left(\begin{array}{ccc}1 & \epsilon_1 & \epsilon_2\\-\epsilon_1 & \epsilon_3 & 1\\-\epsilon_2& 1 &  \epsilon_4\end{array}\right)~.
\label{VRa}
 \end{eqnarray}
This form of $V_R$ is motivated by the need to generate $\overline{c}_R \gamma_\mu b_R W_R^\mu$ coupling.  Here $|\epsilon_i| \ll 1$ are small parameters needed only for cosmology. For collider phenomenology we could set $\epsilon_i$ to zero, but in this case there would be additional symmetries which would make some of the vector-like quarks absolutely stable.  Tiny values of $\epsilon_i \sim 10^{-6}$ would lead to their decay at cosmologically acceptable time scales~\cite{DMZ}.

 With this choice of Yukawa coupling and mass matrices, after rotating the fields to remove $V_L^\dagger$ and $V_R^\dagger$ in Eq. (\ref{flip}) so that they appear in the $W_L^\pm$ and $W_R^\pm$ interactions, the quark mass matrices become diagonal except in the $c-C$  and the $b-B$ sectors, where they are given by the matrices of the seesaw form:
 \begin{eqnarray}
 {\cal M}_{c-C} = \left(\begin{matrix}0 & Y_2^u \kappa_L \\ Y_2^{\prime u} \kappa_R & M_2  \end{matrix} \right),~~~
 {\cal M}_{b-B} = \left(\begin{matrix}0 & Y_3^d \kappa_L \\ Y_3^{\prime d} \kappa_R & M_3  \end{matrix} \right)~.
 \label{seesawa}
 \end{eqnarray}
 The light quark masses are then obtained to be
 \begin{eqnarray}
 m_u &=& Y_1^u \kappa_L,~~m_c \simeq \frac{Y_2^u Y_2^{\prime u} \kappa_L \kappa_R}{M_2},~~m_t = Y_3^u \kappa_L \nonumber \\
 m_d  &=& Y_1^d \kappa_L,~~ m_s = Y_2^d \kappa_L,~~m_b \simeq \frac{Y_3^d Y_3^{\prime d} \kappa_L \kappa_R}{M_3}~.
\label{light}
 \end{eqnarray}
 The heavy quark masses, on the other hand, are found to be:
 \begin{eqnarray}
 M_U = Y_1^{\prime u} \kappa_R,~~ M_C \simeq M_2,~~M_T = Y_3^{\prime u} \kappa_R \nonumber \\
 M_D = Y_1^{\prime d} \kappa_R,~~ M_S \simeq Y_2^{\prime d} \kappa_R,~~M_B \simeq M_3~.
 \label{heavy}
 \end{eqnarray}

 We choose the couplings $(Y_1^u, Y_3^u)$ and $(Y_1^d, Y_2^d)$ hierarchically to fit the masses of $(u,t)$ and $(d,s)$ quarks. It is clear from Eqs. (\ref{light})-(\ref{heavy}) that for low values of $\kappa_R \simeq 2$ TeV needed to explain $R(D^*,D)$ anomaly, ($Y_1^{\prime u}, Y_3^{\prime u}$) cannot be equal to $(Y_1^u, Y_3^u)$ -- as that would lead to light vector-like fermions excluded by the LHC -- and similarly for the down quark sector.  Hence the need to assume parity violation in this type of flavor choice.

The zeros in $M_F$ in Eq. (\ref{flip}) implies that $W_R$ couples only to the heavy quarks $(U,T,D,S)$ and not to the corresponding light quarks $(u,t,d,s)$. On the other hand, for the bottom and charm quarks, the masses are given by the quark seesaw formula and therefore these light fields have both $W_L$ and $W_R$ interactions. The $W_R^\pm$ couplings to the physical quark fields is given by
\begin{eqnarray}
{\cal L}^q_{W_R^\pm} = \frac{g_R}{\sqrt{2}} \left(\overline{U}_R~ \overline{c}_R~ \overline{T}_R\right) \gamma_\mu V_R \left(\begin{matrix}D_R \\ S_R \\ b_R  \end{matrix} \right)\,W_R^{+ \mu} + h.c.
\label{ccq}
\end{eqnarray}
With the form of $V_R$ given in Eq. (\ref{VRa}),
this interaction clearly contains the desired term $\overline{c}_R \gamma_\mu b_R W_R^{+ \mu}$ for explaining the $R(D^*,D)$ anomaly, and no other term involving the light quarks that could lead to unacceptable flavor violation.

In the charged lepton sector we choose for the matrix ${\cal M}_E$ in Eq. (\ref{mass})
\begin{eqnarray}
Y_E = {\rm diag}(Y_1^e, Y_2^e, Y_3^e), ~~Y_E' = {\rm diag}(Y_1^{\prime e}, Y_2^{\prime e}, Y_3^{\prime e}), ~~M_E = {\rm diag}(0,0, M_E)~.
\end{eqnarray}
This leads to decoupled $e$ and $\mu$ fields, while the $\tau$ lepton mixes with the $E_3$ field via the seesaw mass matrix
\begin{eqnarray}
 {\cal M}_{\tau-E_3} = \left(\begin{matrix}0 & Y_3^e \kappa_L \\ Y_3^{\prime e} \kappa_R & M_E  \end{matrix} \right)~.
\end{eqnarray}
The light and heavy lepton masses are then given by
\begin{eqnarray}
m_e &=& Y_1^e \kappa_L,~~m_\mu = Y_2^e \kappa_L,~~m_\tau \simeq \frac{Y_3^e Y_3^{\prime e} \kappa_L \kappa_R}{M_E} \nonumber \\
M_{E_1} &=& Y_1^{\prime e} \kappa_R,~~M_{E_2} = Y_2^{\prime e} \kappa_R,~~M_{E_3} \simeq M_E~.
\label{heavylep}
\end{eqnarray}
This structure leads to the leptonic interactions of $W_R$ given by
\begin{eqnarray}
{\cal L}^{\ell} _{W_R^\pm} = \frac{g_R}{\sqrt{2}} \left(\overline{E}_{1R}~ \overline{E}_{2R}~ \overline{\tau}_R\right) \gamma_\mu \left(\begin{matrix}\nu_{e_R} \\ \nu_{\mu_R} \\ \nu_{\tau_R}  \end{matrix} \right)\,W_R^{- \mu} + h.c.
\label{ccl}
\end{eqnarray}
We see that the only interactions of $W_R$ with light leptons is of the form $\overline{\tau}_R \gamma_\mu \nu_{\tau_R} W_R^{- \mu}$, which is the desired coupling to explain $R(D^*,D)$. Integrating out the $W_R$ field using Eq. (\ref{ccq}) and Eq. (\ref{ccl}) would induce a unique effective dimension six operator involving light quarks and leptons given by ${\cal H}_{eff}\simeq \frac{g^2_R}{2M^2_{W_R}}\bar b_R\gamma_\mu c_R\bar{\nu}_{\tau_R}\gamma^\mu{\tau_R} + h.c.$
Its contribution to $R(D^*,D)$ will be analyzed in Sec. 5.

\subsection{ Avoiding flavor changing neutral current constraints}

As is well known, the right-handed $W_R$ interactions contribute to flavor changing effects such as to $K_L-K_S$, $B_s-\bar{B}_s$ and $B_d-\bar{B}_d$ mixings at the one loop level via box diagrams.  The dominant new contributions arise from the  $W_L-W_R$ mediated box graphs~\cite{FCNC}. In the context of LR models without vector-like quarks, such constraints put $W_R$ mass to be $M_{W_R}/g_R \geq 2.5$ TeV, assuming that the left-handed CKM mixing matrix $V_L$ and its right-handed counterpart $V_R$ are equal. For $g_R \simeq 2$, which is what would be needed to explain $R(D^*,D)$ anomaly, the limit on $W_R$ mass is of order 5 TeV, much above the needed value to explain $R(D^*,D)$.  In this subsection, we show how the flavor structure for the Yukawa couplings and the mass matrices shown in Eqs. (\ref{flip})-(\ref{seesawa}) completely evades these bounds.

\begin{figure}[htb]
\begin{center}
\includegraphics[width=6.cm]{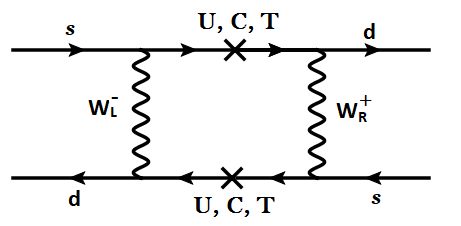}
\end{center}
\caption{Dominant diagrams inducing $\Delta F = 2$ interactions such as $K^0-\bar{K^0}$ mixing in the LR parity asymmetric quark seesaw model.}
\label{box0}
\end{figure}

In the parity asymmetric version of the quark seesaw model, the dominant contributions to $\Delta F=2$ flavor changing effects arise from diagrams of such as the one shown in Fig. \ref{box0}. These amplitudes can be  symbolically written as follows:
\begin{eqnarray}
\Delta M_K\propto (V_{L})_{is} (M_{U})_{ij} (V_{R})_{jd}^* (V_{R})_{\ell s} (M_{U})_{k \ell} (V_{L})_{kd}^*\nonumber\\
\Delta M_{B_s}\propto (V_{L})_{is} (M_{U})_{ij} (V_{R})_{jb}^* (V_{R})_{\ell s} (M_{U})_{k \ell} (V_{L})_{kb}^*\nonumber\\
\Delta M_{B_d}\propto (V_{L})_{ib} (M_{U})_{ij} (V_{R})_{jd}^* (V_{R})_{\ell b} (M_{U})_{k \ell} (V_{L})_{kd}^*\nonumber\\\
\Delta M_{D}\propto V_{L})^*_{ci} (M_{D})_{ij} (V_{R})_{uj} (V_{R})^*_{c \ell} (M_{D})_{k \ell} (V_{L})_{uk}~.
\end{eqnarray}
Note that by our choice of matrices, $M_U V_R=O(\epsilon)$ where $\epsilon$ can be a very small number, of order $10^{-6}$ or so.
This removes the $K$ and  $B_{d,s}$ meson mixing constraints from the dominant source. Furthermore, since  $(M_DV_R^T)_{iu}=O(\epsilon)$ this also removes the $D-\bar{D}$ mixing constraint.

The absence of new contributions to $K-\overline{K}$, $B_{d,s}-\overline{B}_{d,s}$ and $D-\overline{D}$ mixing in the model can also be seen directly from the charged current $W_R$ interaction of Eq. (\ref{ccq}), which is written in terms of physical mass eigenstates, along with the adopted form of $V_R$ of Eq. (\ref{VRa}).  These meson mixing diagrams simply do not connect.

The mixing of $b-B$, $c-C$ and $\tau-E_3$ as given by Eq. (\ref{seesawa}) and Eq. (\ref{ccl}) would imply that there is some amount of flavor violation in the model.  While such mixings in the right-handed sector do not lead to modifications in the interactions of $b,c,\tau$ with the $Z$ boson -- as these right-handed fields have identical SM gauge quantum numbers, the miixings of left-handed fields will modify the interactions of $(b_L, c_L, \tau_L)$ with respect to the standard model.  The charged current $W_L^\pm$ interactions in the physical mass eigenbasis for the quarks is modified and is given by
\begin{eqnarray}
{\cal L}^q_{W_L^\pm} = \frac{g_L}{\sqrt{2}} \left(\overline{u}_L~ \overline{c}_L~ \overline{t}_L ~ \overline{C}_L \right) \gamma_\mu J_L \left(\begin{matrix}d_L \\ s_L \\ b_L \\ B_L  \end{matrix} \right)\,W_L^{+ \mu} + h.c.
\end{eqnarray}
where $J_L$ is given by
\begin{eqnarray}
J_L = \left(\begin{matrix}~V_{ud} & ~V_{us} & V_{ub} c_b & V_{ub} s_b \\ V_{cd} c_c & V_{cs} c_c & V_{cb} c_c c_b & V_{cb} c_c s_b \\
~V_{td} & ~V_{ts} & V_{tb} c_b & V_{tb} s_b \\ V_{cd} s_c & V_{cs} s_c & V_{cb} s_c c_b & V_{cb} s_c s_b       \end{matrix}\right)~.
\label{KL}
\end{eqnarray}
Here $V_{ij}$ stand for elements of the left-handed CKM matrix $V_L$, while $s_c = \sin\theta_c$ and $s_b = \sin\theta_b$ stand for the $c_L-C_L$ and $b_L-B_L$ mixing angles with $c_c = \cos\theta_c$ and $c_b = \cos\theta_b$. These angles are given by (see. Eq. (\ref{seesawa})):
\begin{equation}
\theta_c \simeq \frac{Y_2^u \kappa_L}{M_2} \simeq \frac{m_c}{Y_2^{\prime u} \kappa_R},~~~~~~ \theta_b \simeq \frac{Y_3^d \kappa_L}{M_3} \simeq \frac{m_b}{Y_3^{\prime d} \kappa_R}
\label{second}
\end{equation}
Eq. (\ref{KL}) takes into account these mixings, and the fact that the gauge eigenstates $C_L$ and $B_L$ have no direct couplings to $W_L$. In the second halves of Eq. (\ref{second}) we made use of the light eigenvalue for $c$ and $b$ quarks given in Eq. (\ref{light}). Note that these mixing angles can be as small as $10^{-3}$, corresponding to $\kappa_R = 1.4$ TeV and $Y_2^{\prime u} \sim Y_3^{\prime d} \sim 1$.

The modification of $W_L^\pm$ interactions with quarks will lead to some flavor violation.  The Standard Model box diagram contributions to $K-\overline{K}$ mixing is now given by \cite{inami}
\begin{eqnarray}
H_{\rm eff}^{LL} = \frac{G_F}{\sqrt{2}} \frac{\alpha}{4\pi \sin^2\theta_W} \lambda_i \lambda_j\left[\left(1 + \frac{x_i x_j}{4}\right) I_2(x_i,x_j,1)-2x_i x_j I_1(x_i,x_j,1)      \right] (\overline{s}_L \gamma_\mu d_L)^2~.
\end{eqnarray}
Here $x_i$, $\lambda_i$ and the loop functions $I_i$ are defined as
\begin{eqnarray}
x_i &=& \frac{m_i^2}{M_{W_L}^2}, ~~~~i=u,c,t,C;~~~~\lambda_i = (J_L)_{is}^* (J_L)_{id} \nonumber \\
I_1(x_i,x_j,\eta) &=& \frac{\eta {\rm ln}(1/\eta)}{(1-\eta)(1-x_i\eta)(1-x_j\eta)} + \frac{x_i {\rm ln}x_i}{(x_i-x_j)(1-x_i)(1-x_i\eta)} + (i \rightarrow j), \nonumber \\
I_2(x_i,x_j,\eta) &=&  \frac{{\rm ln}(1/\eta)}{(1-\eta)(1-x_i\eta)(1-x_j\eta)} + \frac{x_i^2 {\rm ln}x_i}{(x_i-x_j)(1-x_i)(1-x_i\eta)} + (i \rightarrow j) \nonumber \\
I_1(x_i,x_j,1) &=& {\rm lim}_{\eta \rightarrow 1} I_1(x_i,x_j,\eta)~.
\label{I12}
\end{eqnarray}
Similar expressions appear in $W_L-W_R$ exchange diagrams, where we shall define $\eta = M_{W_L}^2/M_{W_R}^2$.
 Analogous expressions can be written down for $B_{d,s}-\overline{B}_{d,s}$ mixing as well as $D-\overline{D}$ mixing by interchanging the flavor indices appropriately.  Since $J_L$ is not unitary,  GIM cancellation is no longer effective.  However, deviations are quite small.  For example, with $M_C = 3$ TeV for the charm partner mass, $K^0-\overline{K^0}$ mixing limit requires $\theta_c \leq 0.03$, which is well within the allowed range of the model.  The constraint on $\theta_c$ from $B_{d,s}-\overline{B}_{d,s}$ mass splitting is also of the same order.

\subsection{Universality and other flavor constraints}

The mixing of $c$-quark with the vector-like $C$-quark, $b$ with $B$ and $\tau$ with $E_3$ would imply some modifications in precision electroweak parameters and universality in leptonic decays.  As we shall see below, our benchmark points needed to explain $R(D^*,D)$ anomaly are fully consistent with these constraints.

Lepton universality will be violated owing to $\tau_L-E_{3L}$ mixing.  In the charged current interactions of $W_L^\pm$, this mixing will introduce a factor of $\cos\theta_\tau$ wherever $\tau_L$ appears, which would lead to the modified interaction
\begin{equation}
{\cal L}_\tau^{W_L} = \frac{g_L}{\sqrt{2}} \cos\theta_\tau \overline{\tau}_L \gamma^\mu \nu_{\tau L} W_L^- + h.c.
\end{equation}
The decay $\tau \rightarrow \pi + \nu_\tau$ will be modified, in relation to $\pi \rightarrow \mu \nu_\mu$.  The ratio of the effective couplings, $A_\pi = G_{\tau \pi}^2/G_F^2$ provides the following constraint ($s_\tau = \sin\theta_\tau$):
\begin{equation}
A_\pi = \frac{G_{\tau \pi}^2}{G_F^2}
 = 1- s_\tau^2 = 1.0020 \pm 0.0073\, \cite{PDG}.
\end{equation}
Using 1 sigma error, this would lead to the bound $s_\tau \leq 0.073$.  This constraint, while nontrivial, is easily satisfied within the model, where $s_\tau$ is allowed to be as small as $0.001$.

The interactions of the $Z$ boson with light fermions are modified because of their mixings with vector-like fermions.  However, in the right-handed fermion sector, there are no modifications, as the vector-like fermions have the same SM quantum numbers as the usual fermions.  The interactions of $Z$ with light fermions are then modified to
\begin{equation}
{\cal L}^Z = \frac{g}{2 c_W} \left[\overline{f}_L \left\{T_{3L}^f (1-s_f^2)-Q_f s_W^2\right\} \gamma^\mu f_L + \overline{f}_R (-Q_f s_W^2) \gamma^\mu f_R\right]Z^\mu,
\label{Znew}
\end{equation}
where $s_f$ denotes the mixing of the left-handed fermion $f_L$ with a vector-like fermion.  Here $s_W = \sin\theta_W$ is the weak mixing angle, and $c_W= \cos\theta_W$.  The polarization asymmetry parameter $A_f$, measured at LEP and SLD from forward-backward asymmetry and left-right asymmetry, is now modified to
\begin{equation}
A_f = A_f^{\rm SM} \left(1+ \frac{\delta A_f}{A_f^{\rm SM}}  \right)
\end{equation}
where
\begin{equation}
\frac{\delta A_f}{A_f^{\rm SM}} \simeq \frac{-4 \, Q_f^2 \, s_W^4 \, s_f^2\, \{T_{3L}^f-Q_f\, s_W^2\}}{\{T_{3L}^f-2 \,Q_f\, s_W^2\} \{(T_{3L}^{f})^2-2\, Q_f\, s_W^2\, T_{3L}^f + 2\, Q_f^2\, s_W^4\}}~.
\label{delta}
\end{equation}
Eq. (\ref{delta}), when applied to $c$ and $b$ quarks and $\tau$ lepton would lead to the following shifts (using $s_W^2 = 0.2315$):
\begin{eqnarray}
\frac{\delta A_b}{A_b^{\rm SM}} = -0.158 \,s_b^2,~~~\frac{\delta A_c}{A_c^{\rm SM}} = -1.20 \,s_c^2,~~~\frac{\delta A_\tau}{A_\tau^{\rm SM}} = -12.38\, s_\tau^2~.
\end{eqnarray}
Using experimental values of $A_b$, $A_c$ and $A_\tau$, which are given by \cite{PDG} $A_b = 0.923 \pm 0.020$, $A_c = 0.670 \pm 0.027$ and $A_\tau = 0.1439 \pm 0.0043$ (from $Z$ pole data at LEP), and the theoretical values based on SM given by $A_b^{\rm SM} = 0.9347$, $A_c^{\rm SM} = 0.6677$ with negligible errors, and $A_\tau = A_\ell = 0.1469$, we obtain with 1 sigma error allowance the following limits on the mixing angles:
\begin{equation}
s_b \leq 0.463,~~~ s_c \leq 0.176,~~~s_\tau \leq 0.048~.
\end{equation}
If we use the SLD value of  $A_\tau = 0.136 \pm 0.015$ instead \cite{PDG}, which is somewhat discrepant from the LEP value, we would get $s_\tau \leq 0.091$.

The partial decay widths of the $Z$ boson into $b \overline{b}$, $c \overline{c}$ and $\tau^+ \tau^-$ will deviate from their SM values by an amount given by
\begin{equation}
\frac{\Delta \Gamma_f}{\Gamma_f^{\rm SM}} = \frac{2 Q_f s_W^2 T_{3L}^f s_f^2}{(T_{3L}^f - Q_f s_W^2)^2+(Q_f s_W^2)^2}~,
\end{equation}
leading to
\begin{equation}
\frac{\Delta \Gamma_b}{\Gamma_b^{\rm SM}} = 0.418 s_b^2,~~~\frac{\Delta \Gamma_c}{\Gamma_c^{\rm SM}} = 1.077 s_c^2,~~~\frac{\Delta \Gamma_\tau}{\Gamma_\tau^{\rm SM}} = 1.840 s_\tau^2~.
\end{equation}
The ratio $\Gamma(Z \rightarrow \tau \tau)/\Gamma(Z \rightarrow ee) = 1.0019 \pm 0.0032$ is well measured experimentally.  Compared to the SM, this ratio is modified by a factor $1- s_\tau^2$.  Using 1 sigma error, we find a limit
\begin{equation}
s_\tau \leq 0.053~.
\end{equation}
Similarly, $R_b = \Gamma(Z \rightarrow bb)/\Gamma(Z \rightarrow {\rm hadrons})$ is modified from its SM value to $R_b = R_b^{\rm SM}(1+ 0.418 s_b^2)$.  From the experimental value of $R_b = 0.21629 \pm 0.00066$ \cite{PDG}, we obtain a limit
\begin{equation}
s_b \leq 0.085~.
\end{equation}
A similarly defined ratio $R_c$ is modified to $R_c = R_c^{\rm SM}(1+1.077 s_c^2)$.  Comparing with the experimental value $R_c = 0.1721 \pm 0.0030$ we obtain
\begin{equation}
s_c \leq 0.127~.
\end{equation}

All these constraints are seen to be consistent with the model parameters required to explain $R(D^*,D)$.
We thus conclude that the model in its parity asymmetric form can lead to the desired flavor structure of $W_R^\pm$ currents without inducing unwanted flavor violation in other sectors. In Sec. 5 we show how this flavor structure enables us to explain $R(D^*,D)$ in terms of right-handed currents.
Most of the constraints derived and found to be satisfied in this section also apply to the parity symmetric scenario discussed in the next section.

\section{Parity symmetric flavor structure without FCNC}

In this section we develop a scenario which explains $R(D^*,D)$ anomaly via right-handed currents that is also Parity symmetric.  Apart from its aesthetic appeal, such a scheme can also solve the strong CP problem using Parity symmetry without the need for an axion \cite{bm}.  Our setup is identical to the one discussed in the previous section, with the gauge symmetry being $SU(3)_c \times SU(2)_L \times SU(2)_R \times U(1)_{B-L}$.  This version of the LR model has been shown to solve the strong CP problem owing to the structure of the quark mass matrices that is Parity invariant.  First of all, Parity sets $\theta_{QCD}$ to zero.  Under $P$, fermions transform as $q_L \leftrightarrow q_R$, $\psi_L  \leftrightarrow \psi_R$, $(U,D,E)_L \leftrightarrow (U,D,E)_R$, while the Higgs fields transform as $\chi_L \leftrightarrow \chi_R$. Simultaneously the gauge fields transform as $W_L \leftrightarrow W_R$.  Consequently, the seesaw quark mass matrices take the form
\begin{eqnarray}
{\cal M}_{U,D} = \left( \begin{matrix} 0 & Y_{U,D} \, \kappa_L \\ Y_{U,D}^\dagger \, \kappa_R & M_{U,D}  \end{matrix}\right)~
\label{mat}
\end{eqnarray}
with the condition $M_{U,D}^\dagger = M_{U,D}$.  By separate $SU(2)_L$ and $SU(2)_R$ gauge rotations the VEVs of $\chi_L$ and $\chi_R$, $\kappa_L$ and $\kappa_R$, can be chosen to be real.  The determinant of ${\cal M}_U . {\cal M}_D$ is then real, implying that $\overline{\theta} = 0$ at tree level.  It has been shown that in this setup, there is no induced $\overline{\theta}$ at one loop level \cite{bm}.  We shall briefly review this result in this section, where we show that soft breaking of $P$ which allows for $M_{U,D} \neq M_{U,D}^\dagger$ does not spoil this result. A small value of $\overline{\theta}$ is induced via two loop diagrams, estimated to be $\overline{\theta} \sim 10^{-10}$, which is consistent, but not very far from the limit obtained from neutron electric dipole moment \cite{bm,hall}.

It will be desirable to keep the solution to the strong CP problem of the setup and at the same time provide an explanation for the $R(D^*,D)$ anomaly.  This is what we take up in this section.  Parity was explicitly broken in the discussion of Sec. 3, which therefore has no relevance to the strong CP solution.  Recall that a flipping of $u_R$ and $U_{R}$ (and similarly other quark and lepton fields) played an important role in the discussion of Sec. 3.  The bare mass terms for certain vector-like quarks were set to zero to achieve such flips.  Parity can then not be imposed, or else the masses of the $u$ and $U$ quarks will be in the ratio $\kappa_L/\kappa_R$ which should be of order $(1/10-1/20)$ in order to explain $R(D^*,D)$.  The resulting light vector-like quarks are not allowed by experimental limits.

In the up-quark sector, consider the case where $Y_U$ is proportional to the identity matrix, and $M_U$  an arbitrary non-hermitian matrix:
 \begin{eqnarray}
Y_U &=& y_u \times {\rm diag}(1,1,1),~~~  M_U = V_R^0.\, {\rm diag}({M_1^u, M_2^u, M_3^u}).\,V_L^{0 \dagger} ~.
\label{YU}
\end{eqnarray}
One can remove the unitary matrices $V_L^0$ and $V_R^0$ appearing In Eq. (\ref{YU}) by the following field transformations:
\begin{eqnarray}
U_L =V_R^0 U_L^0,~~~U_R=V_L^0 U_R^0,~~~u_R = V_R^0 u_R^0,~~~u_L = V_L^0 u_L^0 ~.
\end{eqnarray}
This will induce a flavor structure $V_L^0$ in the $W_L$ and $V_R^0$ in the $W_R$ charged current interactions, with $V_L^0$ and 
$V_R^0$ approximately -- but not exactly -- being the left-handed and right-handed CKM matrices.  Note that $V_L^0$ and $V_R^0$ are unrelated. In the new basis, the up-quark mass matrix becomes block-diagonal, with each block given by
\begin{eqnarray}
{\cal M}_{u_i}= 
\left( \begin{matrix}  0 & y_u \kappa_L \\ y_u \kappa_R & M_i^u \end{matrix} \right)~. 
\end{eqnarray}
For the up and charm quarks, with $M_ i^u \gg y_u \kappa_R$, the eigenvalues are given as
\begin{eqnarray}
m_u &\simeq& \frac{y_u^2 \kappa_L \kappa_R}{M_1^u}, ~~~M_U \simeq M_1^u \nonumber \\
m_c &\simeq&  \frac{y_u^2 \kappa_L \kappa_R}{M_2^u}, ~~~M_C = M_2^u~.
\end{eqnarray}
As for the top quark, the $t-T$ mixing in the right-handed sector cannot be too small, and hence the seesaw formula that applies to
$u$ and $c$ quarks is not applicable.  The reason is that $M_T \equiv M_3^u$ cannot be taken to be much larger than $y_u \kappa_R$, or else the top quark mass will be suppressed compared to the electroweak scale $\kappa_L$.  The physical top quark state and its partner $T$ quark state are given as ($c_t = \cos\theta_t$, $s_t = \sin\theta_t$, $t^0$ and $T^0$ are mass eigenstates)
\begin{equation}
t_R^0 = c_t t_R + s_t T_R,~~~T_R^0 = - s_t t_R + c_t T_R~
\end{equation}
with the $t_R-T_R$ mixing angle given as
\begin{equation}
\tan\theta_t = \frac{y_u \kappa_R}{M_3^u}~.
\end{equation}
Analogous mixing in the $t_L-T_L$ sector is small, given by replacing $\kappa_R$ by $\kappa_L$ in $\tan\theta_t$.

We shall take  the limit $M_3^u \ll  y_u \kappa_R$, so that the mass eigenvalues are:
\begin{equation}
m_t \simeq y_u \kappa_L,~~~M_T \simeq y_u \kappa_R ~.\nonumber
\end{equation}
This corresponds to a flip of $t_R \leftrightarrow T_R$, implying that the light top $t_R$ will not have $W_R$ interactions. Such a choice, with $c_t \rightarrow 0$,  helps with suppressing FCNC arising from $W_L-W_R$ mixed box diagrams, as discussed later.

In the down quark mass matrix, the  matrices ${Y}_D$ and ${M}_D$ of Eq. (\ref{mat}) are chosen as
\begin{eqnarray}
{Y}_D = \left( \begin{matrix}  0 & Y_1^d & 0 \\ Y_2^d & 0 & 0 \\ 0 & 0 & Y_3^d \end{matrix} \right) ,~~ ~~
{M}_D = \left( \begin{matrix}  0 & M_1^d & 0 \\ M_2^d & 0 & 0 \\ 0 & 0 & M_3^d \end{matrix} \right)  ~.
\end{eqnarray}
This mass matrix consists of three $2 \times 2$  block-diagonal matrices:
\begin{eqnarray}
{\cal L}_{\rm mass}^d &=&\left( \begin{matrix} \overline{d}_{1L} & \overline{D}_{1L}  \end{matrix} \right) \left( \begin{matrix} Y_1^d \kappa_L & 0 \\ M_1^d & Y_2^d \kappa_R  \end{matrix}  \right) \left( \begin{matrix} D_{2R} \\ d_{2R}   \end{matrix}  \right) + \left( \begin{matrix} \overline{d}_{2L} & \overline{D}_{2L}  \end{matrix} \right) \left( \begin{matrix} 0 & Y_2^d \kappa_L  \\  Y_1^d \kappa_R & M_2^d \end{matrix}  \right) \left( \begin{matrix} d_{1R} \\ D_{1R}   \end{matrix}  \right) \nonumber \\
&+& \left( \begin{matrix} \overline{d}_{3L} & \overline{D}_{3L}  \end{matrix} \right) \left( \begin{matrix}  0 &Y_3^d \kappa_L \\ Y_3^d \kappa_R & M_3^d  \end{matrix}  \right) \left( \begin{matrix} d_{3R} \\ D_{3R}   \end{matrix}  \right)  + h.c.
\label{block}
\end{eqnarray}
The third block is the usual seesaw matrix, identified as the $b-B$ sector. The eigenvalues are given approximately by
\begin{equation}
m_b \simeq \frac{(Y_3^d)^2 \kappa_L \kappa_R}{M_3^d},~~~m_B \simeq M_3^d ~.
\end{equation}

The first block in Eq. (\ref{block}) turns out to be the $s-S$ sector.  
We take  $M_1^d \sim Y_2^d \kappa_R$ in this block,  so that $d_{2R}-D_{2R}$ mixing is significant. We shall further take  the limit $M_1^d \rightarrow 0$, in which case the light state will be composed of $D_{2R}$ with the $d_{2R}$ belonging to the heavy state.  Analogous to the $t-T$ sector, we identify the physical states as
\begin{equation}
s_R^0 = c_s s_R + s_s S_R,~~~S_R^0 = - s_s s_R + c_s S_R~
\end{equation}
with the $s_R-S_R$ mixing angle given as
\begin{equation}
\tan\theta_s = \frac{Y_2^d \kappa_R}{M_1^d}~.
\end{equation}
The eigenvalues $m_s$ and $m_S$  of the first block matrix are:
\begin{equation}
m_s \simeq Y_1^d \kappa_L,~~~ M_S \simeq Y_2^d \kappa_R.
\label{ms}
\end{equation}
Note that this flips $d_{2R}$ with $D_{2R}$.  
 That is, $d_{2R}$ is the heavy state that couples to $W_R$ while $D_{2R}$ is the light state with no coupling to $W_R$.

For the second block matrix in Eq. (\ref{block}), we take  $M_2^d \gg Y_1^d \kappa_R$, leading to the eigenvalues:
\begin{equation}
m_d \simeq \frac{Y_1^d Y_2^d \kappa_L \kappa_R}{M_2^d,}~~~ m_D \simeq M_2^d~. \nonumber
\end{equation}
This ligther eigenvalue is smaller than the lighter eigenvalue of the first block, $m_s \simeq Y_1^d \kappa_L$, see Eq. (\ref{ms}), and therefore should be identified as the $d$-quark.  Thus, $d_R$ couples to $W_R$.

The identification  in the limit $\cos\theta_s \rightarrow 0$ is this: $d_{1R}= d_R$, $d_{2L} = d_L$, $d_{1L} = s_L$, $D_{2R} = s_R$.  The flip $d_{2L} \leftrightarrow 
d_{1L}$ is no concern, since that can be compensated by the arbitrary form of $V_L^0$ in $W_L$ charged current.  In fact, with this interchange implemented, $V_L^0$ will be identified as the left-handed CKM matrix $V_L$.

Suppose the form of $V_R^0$ in Eq. (\ref{YU}) is
\begin{eqnarray}
V_R^0 = \left( \begin{matrix} 0 & 1 & 0 \\ 0 & 0 & 1 \\ 1 & 0 & 0   \end{matrix} \right)~.
\label{VR0}
\end{eqnarray}
This form of $V_R^0$ is motivated by maximizing new contributions to $R(D^*,D)$ -- with the (2,3) entry being 1.  The flippling $s_R \leftrightarrow S_R$ helps with suppressing the decay $\tau \rightarrow K \nu_\tau$ which would set significant constraints on new contributions to $R(D^*,D)$, if it is allowed.
The $\overline{u}_R \gamma^\mu d_{2R} W_R^\mu$ coupling will now involve the heavy $D_{2R}$ state and will not lead to $\tau \rightarrow K \nu$ decay.

This form of the right-handed CKM matrix $V_R^0$ is chosen to fit the $R(D^*,D)$ anomaly via right-handed currents while suppressing new contributions to $K^0-\overline{K^0}$, $B_{d,s}-\overline{B}_{d,s}$ and $D^0-\overline{D^0}$ mixing mediated by $W_L-W_R$ mixed box diagrams.  The amplitude for such mixed box diagrams, while suppressed by a factor of $(g_R^2/g_L^2)(M_{W_L}^2/M_{W_R}^2)$, is enhanced by a numerical factor of about $10^3$ arising from combinatorial factor of 8, enhanced matrix element (for the case of $K^0-\overline{K^0}$ mixing) of order 20 and a factor ln$(m_c^2/M_{W_R}^2) \simeq 8$ \cite{FCNC,tran}. Thus, suppression of these mixed box diagram contributions is essential for explaining $R(D^*,D)$ anomaly.  In addition to the form of $V_R^0$ given in Eq. (\ref{VR0}), a second form can also be considered in principle, with the interchange of first and second column in Eq. (\ref{VR0}).  However, this case, while being consistent with FCNC induced by box diagrams, would lead to to the decay $\tau \rightarrow \pi \nu_\tau$, leading to universality violation at such a level as to make new contributions to $R(D^*,D)$ not significant.  We shall not consider such a form as a result. With these form of $V_R^0$ of Eq. (\ref{VR0}), constraints from $K^0-\overline{K^0}$, $B_{d,s}-\overline{B}_{d,s}$ mixing and $D^0-\overline{D^0}$ mixing can be readily satisfied, as we shall see.  Such a form of $V_R^0$ would lead to excessive meson mixing in the standard formulation of left-right symmetric models, but not in the quark seesaw version.

Including the large $t_R-T_R$ mixing as well as $s_R-S_R$ mixing, the right-handed CKM matrix given in Eq. (\ref{VR0}) appears in the charged current interactions as
\begin{eqnarray}
{\cal L}_{W_R} &=& \frac{g_R} {\sqrt{2}}\left(\overline{u^0_R}, ~\overline{c_R^0},~ \overline{t_R^0}, ~\overline{T_R^0} \right) \left(\begin{matrix} 0 & c_s & 0 & -s_s\\ 0 & 0 & 1 & 0 \\ c_t & 0 & 0  & 0 \\ -s_t & 0 & 0 & 0  \end{matrix}  \right)\gamma^\mu \left(\begin{matrix}d_R^0 \\ s_R^0 \\ b_R^0 \\ S_R^0  \end{matrix}  \right)W_R^{+\mu}+ h.c.
\label{WR12}
\end{eqnarray}
The $4 \times 4$ mixing matrix appearing in Eq. (\ref{WR12}) will be denoted as $V_R$.  
Unlike the light quark partners, the top-quark partner (and the strange quark partner) have to be relatively light.  
Note that there is no light vector-like fermion even with $M_3^u = 0$,  since the Yukawa coupling $y_u$ is of order one.  However, this choice would predict $M_T/m_t = \kappa_R/\kappa_L$, which for explaining $R(D^*,D)$ anomaly is about $10-20$. Thus, the mass of the top partner is in the range
$(1.5-2.5)$ TeV in this scenario. The mixing angle $\theta_t \rightarrow \pi/2$ in this limit, which means that $\cos\theta_t \rightarrow 0$.  All entries in the third row of the $4 \times 4$ matrix $V_R$ in Eq. (\ref{WR12}) vanish for this choice.  Similarly, in the limit $c_s \rightarrow 0$, all entries in the second column of $V_R$ in Eq. (\ref{WR12}) would vanish.  As already noted, this would prevent the decay $\tau \rightarrow K \nu_\tau$.  As a result of $c_t \rightarrow 0$, box diagrams involving $W_L-W_R$ exchange would be suppressed, thus evading stringent flavor constrains from meson-antimeson oscillations.

\begin{figure}[htb]
\begin{center}
\includegraphics[width=5.7cm]{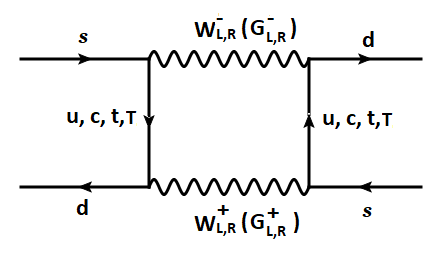}~~~
\includegraphics[width=5.5cm]{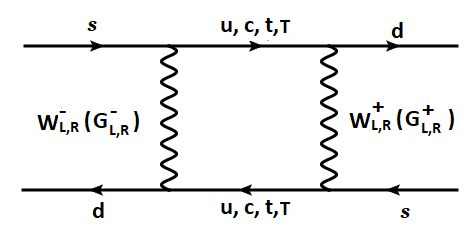}
\end{center}
\caption{Leading $W_L-W_R$ exchange diagram contribution to $K^0-\overline{K^0}$ mass splitting in the parity symmetric LR model.}
\label{box}
\end{figure}

To see the suppression of $W_L-W_R$ box diagrams shown for this case in Fig. \ref{box}, we note that their amplitudes  are given as in standard LR models, but with internal $T$ quark included. The effective Hamiltonian for  $K^0-\overline{K^0}$ mixing is given by \cite{tran}
\begin{eqnarray}
H_{\rm efff}^{LR} &=& \frac{G_F}{\sqrt{2}} \frac{\alpha}{4\pi s_W^2} \lambda_i \lambda_j 2\eta (x_i x_j)^{1/2} \left[(4+x_i x_j \eta)I_1(x_i,x_j,\eta)-(1+\eta)I_2(x_i,x_j,\eta)  \right](\overline{s}_R d_L)(\overline{s}_L d_R) \nonumber \\
\end{eqnarray}
where $\eta = M_{W_L}^2/M_{W_R}^2$, $x_i = m_i^2/M_{W_L}^2$ for $i=u,c,t,T$ and the functions $I_1$ and $I_2$ are defined in Eq. (\ref{I12}). The parameter $\lambda_i$ are defined as
\begin{equation}
\lambda_i \equiv (V_L)_{is}^* (V_R)_{id}~.
\end{equation}
With $V_R$ given by the $4 \times 4$ matrix of Eq. (\ref{WR12}), and with $c_t \rightarrow 0$, the new contributions to $K^0-\overline{K^0}$ mixing vanishes.  The $W_L-W_R$ box diagram would require chirality flips on the $T$-quark internal lines.  However, $T_L$ has no coupling to $W_L$, being a singlet of $SU(2)_L$, and thus there is no contribution to $K^0-\overline{K^0}$ mixing.  Contributions to $B_d-\overline{B}_d$ mixing also vanishes, being proportional to $m_c M_T$.  
New contributions to $B_s-\overline{B}_s$ mixing also vanish, as the second column of $V_R$ is all zero in the limit $c_s \rightarrow0$.  Similarly, new contributions to $D^0-\overline{D^0}$ also vanishes, since this requires chirality flip of $S$ quark.  
 Thus, the flavor structure in the quark sector is consistent with the most stringent constraints from FCNC.

It should be pointed out that the form of the right-handed CKM matrix given in Eq. (\ref{VR0}) can also be realized in the standard left-right symmetric models without parity symmetry.  There is a related possibility where the first and second column of Eq. (\ref{VR0}) are interchanged.  Such models cannot explain $R(D^*,D)$ anomaly, however. In this case, if we adopt a form  for $V_R^0$ where first and second column are interchanged in Eq. (\ref{VR0}), there would be a new contributions to $B_s-\overline{B_s}$ which goes as $[(V_L)_{tb}^*(V_R)_{ts}]^2 m_t$. We find the constraint from this mixing on the $W_R$ mass to be 40 TeV with $g_R=g_L$, and even stronger if $g_R > g_L$.
Similarly, with form of $V_R^0$ as it is, $K^0-\overline{K^0}$ mixing will receive a contribution proportional to $[(V_L)_{ts}^* (V_R)_{td}]^2 m_t^2$, which leads to a constraint $M_{W_R} \geq 70$ TeV for $g_R = g_L$.  New contributions to $B_d-\overline{B}_d$ mixing will go as
$[(V_L)_{tb}^* (V_R)_{td}]^2 m_t^2$. We obtained a stringent limit of $W_{W_R} \geq 225$ TeV in this case.\footnote{In all cases, when the $W_L-W_R$ diagram gives nonzero contributions, we have followed the matrix element evaluations compiled in the first of Ref. \cite{Zhang} to obtain limits quoted here.}
It is clear that these constraints would contradict the $W_R$ mass of order 2 TeV and $g_R = 2$ needed to explain $R(D^*,D)$.  These contributions are absent in the $P$ symmetric universal seesaw model, when $c_t$ and $c_s$ in Eq. (\ref{WR12}) are small. The standard LR models also does not allow for a suppressed coupling of $Z_R$ with electron which is needed to be consistent with LEP bounds.

With the form of $W_R^\pm$  interaction given in Eq. (\ref{WR12}), $W_R^\pm$ will not be produced resonantly at hadron colliders nor by  by $u-s$ fusion when $c_s \rightarrow 0$. 
 Interactions of Eq. (\ref{WR12}) are exactly of the right form needed to explain the $R(D^*,D)$ anomaly.  For this purpose we should specify the couplings of $W_R^\pm$ to leptons as well to which we now turn.

In the charged lepton sector the seesaw mass matrix has a form as given in Eq. (\ref{mass}). Here again, as in the quark sector, we shall assume that Parity is softly broken in the bare mass terms of the vector-like $E$ fields. As a result, $M_E$ is not hermitian.  This soft breaking in the leptonic sector will help suppress $Z_R$ coupling to electrons, which is strongly constrained by LEP data. This suppression is achieved by flipping the $e_R$ field with a vector-like lepton field, as discussed below.

Flipping of $e_R$ field with one of the $E_R$ fields can be achieved by the following choice for the block mass matrices $Y_E$ and $M_E$ in Eq. (\ref{mass}):
\begin{eqnarray}
Y_E = \left(\begin{matrix} * & * & Y_1^e \\ * & * & Y_2^e \\ * & * & Y_3^e  \end{matrix}  \right),~~~
M_E = \left(\begin{matrix} M_{11} & * & * \\ * & * & M_{23} \\ * & M_{32} & *  \end{matrix}  \right)
\label{star}
\end{eqnarray}
where a * indicates small entry.  When the * entries are ignored, all three chiral families would be massless.  Thus, the couplings $Y_i^e$ are not constrained by the light lepton masses, and can be of order one.  With all the * entries set to zero, this matrix can be exactly diagonalized by the following basis transformations:
\begin{equation}
\psi_R^0 = U_R \psi_R,~~~\psi_L^0 = U_L \psi_L,
\end{equation}
which reads more explicitly as
\begin{eqnarray}
\left(\begin{matrix}e_{1R}^0 \\ e_{2R}^0 \\ e_{3R}^0 \\ E_{1R}^0 \\ E_{2R}^0 \\ E_{3R}^0     \end{matrix}  \right) =
\left[\begin{matrix}
c_{\alpha_R} c_\theta & c_{\alpha_R} s_\theta c_\phi & c_{\alpha_R} s_\theta s_\phi & 0 & -s_{\alpha_R} & 0 \\
0 & s_\phi & -c_\phi & 0 & 0 & 0 \\
s_\theta & -c_\theta c_\phi & -c_\theta s_\phi & 0 & 0 & 0 \\
0 & 0 & 0 & 1 & 0 & 0\\
0 & 0 & 0 & 0 & 0 & 1 \\
s_{\alpha_R} c_\theta & s_{\alpha_R} s_\theta c_\phi & s_{\alpha_R} s_\theta s_\phi & 0 & c_{\alpha_R} & 0
    \end{matrix}   \right] \left( \begin{matrix} e_{1R} \\ e_{2R} \\ e_{3R} \\ E_{1R} \\ E_{2R} \\ E_{3R}     \end{matrix}   \right)~.
    \label{leptonmass}
\end{eqnarray}
Here $e_i^0$ and $E_i^0$ refer to mass eigenstates. The matrix $U_L$ is obtained from the matrix above by replacing $\alpha_R$ by $\alpha_L$ and by interchanging the fifth and sixth rows.  Here we have defined
\begin{eqnarray}
Y_1^e &=& Y^e \cos\theta,~~Y_2^e = Y^e \sin\theta \cos\phi,~~ Y_3^e = Y^e \sin\theta \sin\phi, \nonumber \\
\tan\alpha_R &=& \frac{\kappa_R Y^e}{M_{32}},~~~\tan\alpha_L = \frac{\kappa_L Y^e}{M_{23}}~.
\end{eqnarray}
In Eq. (\ref{leptonmass}), $c_{\alpha_R} = \cos\alpha_R$, $c_\theta = \cos\theta$, $s_\phi = \sin\phi$ and so forth.  The Lagrangian for the lepton masses read as
\begin{equation}
{\cal L}_{\rm mass}^{\rm lep} = M_{11} \overline{E^0}_{1L} E^0_{1R} + \frac{M_{23}}{c_{\alpha_L}} \overline{E^0}_{2L} E^0_{2R} + \frac{M_{32}}{c_{\alpha_R}} \overline{E^0}_{3L} E^0_{3R} + h.c.
\end{equation}

We see that in this limit, all chiral leptons are massless, even when the Yukawa coupling $Y^e$ is of order one.  Furthermore, the angle $\alpha_R$ can be of order one, while $\alpha_L$ is much smaller.  In the limit $M_{32} \rightarrow 0$, and with $\sin\theta = 0$, $e_{1R}$ and $E_{2,3R}$ will be flipped.  That is, $e_{1R}^0 = -E_{2R}$ and $E_{3R}^0 = e_{1R}$. Note that the mass of $E_3$, which is $M_{32}/c_{\alpha_R} = Y^e \kappa_R$ in this limit, and can be of order TeV. This means that  the mass of the vector like partner of electron is less than about 4.5 TeV. However, if $Y^e$ is of order one, $e_L$ can potentially mix with $E_{2L}$ with the mixing angle given by $Y_1^e \kappa_L/M_{23}$.  From lepton universality, this mixing angle should be $\leq 0.03$ or so, which can be satisfied by choosing $M_{23}$ of order 10 TeV.  Note that if we had imposed Parity on the mass terms, $M_{23} = M_{32}^*$, and this solution for $e_R \leftrightarrow E_{2R}$ flipping will be unavailable.

Once the small entries denoted as * in Eq. (\ref{star}) are included, small masses for $e$, $\mu$ and $\tau$ will be generated.  Care should be taken to ensure that the flipping indeed corresponds to $e_R \rightarrow E_{2R}$ and not $\mu_R \rightarrow E_{2R}$.  There is enough freedom in the model to ensure this condition. In what follows, we shall assume that such $e_R \rightarrow E_{2R}$ flipping has been done.

As for flavor violation, the discussions of Sec. 3 apply to the parity symmetric version as well.  The $b_L-B_L$ mixing angle is given as $\theta_b \simeq m_b/(Y_3^d \kappa_R)$ which can be as small as 0.001, thus satisfying constraints from $R_b$.  Similarly, $s_\tau$, $s_c$, etc.,  can be small enough to satisfy their experimental limits.

As for lepton non-universality in $B$-meson decay, we note that
if $\nu_{eR}$ and $\nu_{\mu R}$ are heavy, then the new decays of $b \rightarrow c \,\ell \,\overline{\nu}_{eR}$
and  $b \rightarrow c \,\ell \,\overline{\nu}_{\mu R }$
will be kinematically forbidden, while
the decay $b \rightarrow c \, \ell \, \overline{\nu}_{\tau R}$ will be allowed provided that $\nu_{\tau R}$ is light (which we assume). This scenario can then explain the $R(D^*,D)$ anomaly.

\subsection{A complete theory with Parity}

A complete theory with Parity symmetry should explain why $g_R \neq g_L$, as needed for the $R(D^*,D)$ anomaly.  This can happen at low energies in a variety of ways.  Parity symmetry may be spontaneously broken (without breaking $SU(2)_R$ symmetry) at a high scale $\Lambda$. This can lead to an asymmetric spectrum under $SU(2)_L$ and $SU(2)_R$ in the energy interval $M_I \leq \mu \leq \Lambda$, explaining why $g_L \neq g_R$ at $M_I$.  The scales $\Lambda$ and $M_I$ may be identified with the GUT scale and an intermediate scale where the asymmetric matter sector acquire their masses.

Alternatively, the full gauge symmetry could be $SU(3)_c \times SU(2)_L \times SU(2)_R \times SU(2)_D\times U(1)_{B-L}$, where all fermion fields are neutral under the $SU(2)_D$.  A self-dual bifundamental Higgs field $\Phi_L(1,2,1,2,0)$ spontaneously breaks $SU(2)_L \times SU(2)_D$ down to its diagonal subgroup $SU(2)_{\rm weak}$, which is identified as the weak interaction gauge symmetry.  This filed is accompanied by a right-handed partner field $\Phi_R(1,1,2,2,0)$, which is assumed to have no vacuum expectation value. Such an embedding would lead to the relation
\begin{equation}
g_w^{-2} = g_L^{-2} + g_D^{-2}~
\end{equation}
where $g_w$ is the weak $SU(2)_L$ gauge coupling and $g_D$ is the ``dark" $SU(2)_D$ gauge coupling.  Even with $g_L = g_R$, one obtains $g_w \neq g_R$ this way, and Parity is maintained above this symmetry breaking scale. If $SU(2)_D$ is broken near the TeV scale, this dark sector can also provide interesting dark matter candidates.

The $\Phi_R(1,1,2,2,0)$ field, which does not acquire a VEV, can be an interesting dark matter candidate.  Its existence is required by parity symmetry.  Once $SU(2)_L \times SU(2)_D$ breaks down to the diagonal $SU(2)_w$ by the VEV of $\Phi_L$, the field $\Phi_R$ will transform under
$SU(3)_c \times SU(2)_w \times SU(2)_R \times U(1)_{B-L}$ as a $(1,2,2,0)$ scalar.  This self-dual field has the quantum numbers of a weak doublet, which turns out to be inert.  Thus, a complete parity embedding leads to a natural inert doublet dark matter model \cite{inert}, which has been widely studied.  It should be remarked that for $\Phi_R$ to be a dark matter candidate, an allowed quartic coupling $\chi_L^\dagger \Phi_L \Phi_R \chi_R$ should be absent, which can be arranged by a discrete symmetry. No other couplings will affect the stability of $\Phi_R$ dark matter.

\subsection{Solving the strong CP problem}

Here we briefly review how Parity symmetry solves the strong CP problem in the universal quark seesaw framework \cite{bm,hall}. We have already noted that parity symmetry sets $\theta_{QCD}$ to zero.  Furthermore, Det(${\cal M}_U . {\cal M_D})$ (see Eq. (\ref{mat})) is real, so that there is no tree-level contribution to $\overline{\theta}$. If $\overline{\theta}$ is induced at the one-loop, it would be typically too large, compared to the experimental limit of $\overline{\theta} \leq 10^{-10}$ arising from neutron electric dipole moment.  This is not an issue in our model, as the one-loop contributions to $\overline{\theta}$ are all zero.  This is true even when parity is softly broken in the bare quark mass matrices $M_{U,D}$ in Eq. (\ref{mat}), as shown in Ref. \cite{bm}.  We shall briefly review this result here.

Following Ref. \cite{bm}, we write the up-quark mass matrix including loop corrections as
\begin{equation}
{\cal M}_U = {\cal M}_U^0(1+C)~.
\end{equation}
Then the contribution of up-type quarks to $\overline{\theta}$ given by
\begin{eqnarray}
\overline{\theta} = {\rm Arg}{\rm Det}(1+C) = {\rm Im}{\rm Tr}(1+C) = {\rm Im}{\rm Tr}\,C_1
\end{eqnarray}
where $C = C_1 + C_2 + ...$ is used as a loop expansion. If the loop corrections to ${\cal M}_U$ is written as
\begin{eqnarray}
\delta {\cal M}_U = \left[ \begin{matrix} \delta M_{LL}^U &  \delta M_{LH}^U \\ \delta M_{HL}^U & \delta M_{HH}^U \end{matrix} \right],
\end{eqnarray}
then $\overline{\theta}$ is given by
\begin{eqnarray}
\overline{\theta} = {\rm Im} {\rm Tr} \left[ - \frac{1}{\kappa_L \kappa_R}\delta M_{LL}^U (Y_U^\dagger)^{-1} M_U Y_U^{-1}
+ \frac{1}{\kappa_L} \delta M_{LH}^U Y_U^{-1} + \frac{1}{\kappa_R} \delta M_{HL}^U (Y_U^\dagger)^{-1}  \right]~.
\label{theta}
\end{eqnarray}
Note that the correction terms $\delta M^U_{HH}$ does not appear in $\overline{\theta}$ at the one-loop level.

\begin{figure}[htb]
\begin{center}
\includegraphics[width=5.0cm]{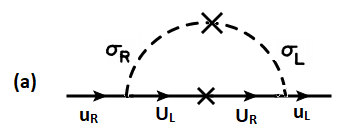}~~~
\includegraphics[width=5.0cm]{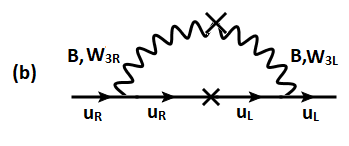}~~~
\includegraphics[width=5.0cm]{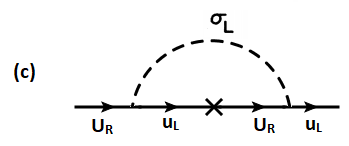}~~~\\
\includegraphics[width=5.0cm]{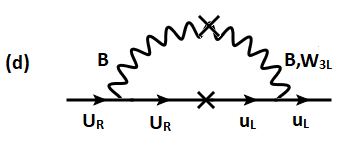}~~~
\includegraphics[width=5.0cm]{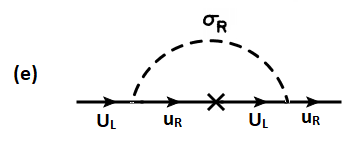}~~~
\includegraphics[width=5.0cm]{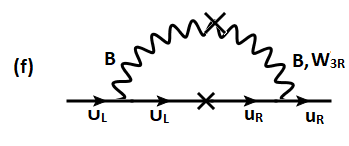}~~~\\
\includegraphics[width=5.0cm]{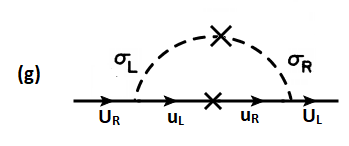}~~~
\includegraphics[width=5.0cm]{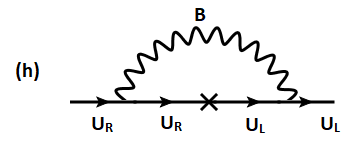}
\end{center}
\caption{One loop corrections to the up quark mass matrix.}
\label{loop}
\end{figure}

The one-loop diagrams that generate corrections to the up-quark mass matrix are shown in Fig. \ref{loop}. In evaluating these diagrams we shall treat the mass matrix as part of the interaction Lagrangian, in which case the cross on the internal fermion line stands for all possible tree-level diagrams where an initial $f_L$ becomes a $f_R$.  Defining $F_{L,R} = (u, U)_{L,R}$, the tree-level mass matrix can be written as $\overline{F}_L {\cal M}_U^0 F_R$ in the Lagrangian.  The full tree-level propagator with all possible mass insertions is then
\begin{eqnarray}
\overline{F}_R \left [{\cal M}_U^{0 \dagger} \frac{k^2}{k^2 - {\cal M}_U^0 {\cal M}_U^{0 \dagger}}   \right] F_L~.
\end{eqnarray}
Now define the inverse matrix
\begin{eqnarray}
\left[{\cal M}_U^{0 \dagger} \frac{k^2}{k^2 - {\cal M}_U^0 {\cal M}_U^{0 \dagger}}- k^2\right]^{-1} = \left[\begin{matrix} X(k^2) & Y(k^2) \\
Y^\dagger(k^2) & Z(k^2)  \end{matrix}  \right]
\end{eqnarray}
with $X = X^\dagger$ and $Z= Z^\dagger$.  Ordinary matrix multiplication determines $X,Y,Z$ as
\begin{eqnarray}
&~&(\kappa_R^2 Y_U^\dagger Y_U + M_U M_U^\dagger - k^2) Y^\dagger = - \kappa_L M_U Y_U^\dagger X \nonumber \\
&~&\kappa_L Y_U Y_U^\dagger X + Y_U M_U^\dagger Y^\dagger = \frac{1}{\kappa_L}(I + k^2 X) \nonumber \\
&~&Y = -\kappa_L H Y_U M_U^\dagger Z
\end{eqnarray}
where
\begin{equation}
H = ( \kappa_L^2 Y_U Y_U^\dagger  - k^2)^{-1} = H^\dagger~.
\end{equation}
The interaction corresponding to the cross on the internal fermion lines of Fig. \ref{loop} can be read off from
\begin{eqnarray}
-{\cal L}_{\rm eff}^{\rm tree} &=& \overline{U}_R \left[\frac{k^4}{\kappa_L}Y_U^{-1} Y(k^2)\right]U_L + \overline{u}_R \left[ k^2 Y_U \kappa_R Z(k^2)\right] U_L \nonumber \\
&+& \overline{U}_R \left[\frac{k^2}{\kappa_L}Y_U^{-1} \{I+ k^2 X(k^2)\}   \right]u_L + \overline{u}_R \left[k^2Y_u \kappa_RY^\dagger(k^2)\right] + h.c.
\end{eqnarray}

Consider the scalar exchange diagram of Fig. \ref{loop} (a).  Its amplitude is given by
\begin{eqnarray}
\delta M_{LL}^U = \int \frac{d^4k}{(2\pi)^4} Y_U \frac{1}{\kappa_L} Y_U^{-1} \frac{k^2 Y(k^2) Y_U^\dagger \lambda_2 \kappa_L \kappa_R}{[(p-k)^2-M_{\sigma_L}^2][(p-k)^2-M_{\sigma_R}^2]}~.
\end{eqnarray}
Its contribution to $\overline{\theta}$, given by Eq. (\ref{theta}), is
\begin{eqnarray}
-{\rm Im}{\rm Tr} \left[\frac{\lambda_2}{\kappa_L} \int \frac{d^4k}{(2\pi)^4}  \frac{k^2 Y(k^2) M_U (Y_U)^{-1}}{[(p-k)^2-M_{\sigma_L}^2][(p-k)^2-M_{\sigma_R}^2]} \right]~.
\end{eqnarray}
We can evaluate the trace before performing the momentum integration, which yields
\begin{equation}
{\rm Tr}[Y(k^2) M_U Y_U^{-1}] = - \kappa_L {\rm Tr} [(Y_U^\dagger Y_U \kappa_L^2 - k^2)^{-1}M_U^\dagger Z(k^2) M_U ]~.
\end{equation}
Since the righ-hand side is the product of two hermitian matrices, its trace is real.  Hence we conclude that the contribution from Fig. \ref{loop} (a) to $\overline{\theta}$ is zero.

The gauge contributions from Fig. \ref{loop} (b) has the same flavor structure as Fig. \ref{loop} (a), viz., $Y(k^2) Y_U^\dagger$.  Therefore, the contribution from Fig. \ref{loop} (b) to $\overline{\theta}$ is also zero.  The off-diagonal contribution from Fig. \ref{loop} (c)-(f) have the matrix structures
\begin{eqnarray}
&~& {\rm Fig.} \ref{loop}~ (c): ~~~\left[I + k^2 X(k^2)  \right]Y_U \nonumber \\
&~& {\rm Fig.} \ref{loop} ~(d): ~~~\left[I + k^2 X(k^2)  \right](Y_U^{\dagger})^{-1}~.
\end{eqnarray}
After multiplying by $Y_U^{-1}$, the relevant trace for $\overline{\theta}$ is found to involve $(I + k^2 X)$ and $(I + k^2 X)(Y_U Y_U^\dagger)^{-1}$. Both these traces are real, since $X$ is hermitian.  Finally, the contribution from Fig. \ref{loop} (e) is proportional to $Y_U^\dagger Y_U Z(k^2)Y_U^\dagger$ and Fig. \ref{loop} (f) is $Z(k^2) Y_U^\dagger$.  These contributions to $\overline{\theta}$ are also vanishing.  Thus we see that all one-loop contributions to $\overline{\theta}$ are zero, even with the bare mass terms $M_{U,D}$ being non-hermitian. There are two-loop diagrams that generate nonzero $\overline{\theta}$, which has been estimate to be of order $10^{-10}$ \cite{bm,hall}, consistent with neutron EDM limits. Thus, this class of LR models provides a solution to the strong CP problem without invoking the axion.

\section{Explaining the \boldmath{$R(D^*,D)$} anomaly}

As mentioned in the introduction, the BaBar, Belle and LHCb collaborations have measured $R(D)$ and $R(D^\ast)$ to very high precision.
The combined experimental values are \cite{HFLAV}:
\begin{eqnarray}
R(D)_{\text{Exp}} &=& 0.407 \pm 0.039 \pm 0.024, \\
R(D^\ast)_{\text{Exp}} &=&  0.306 \pm 0.013 \pm 0.007.
\end{eqnarray}
We see that $R(D)$ and $R(D^*)$ exceed the SM predictions by 2.3$\sigma$ and 3.0$\sigma$ respectively. The net anomaly  is about 3.78$\sigma$.
The SM predictions for $R(D^*)$~\cite{HFLAV} which shows an arithmetic average of theory calculations~\cite{Bernlochner:2017jka, Jaiswal:2017rve, Bigi:2017jbd} is:
 \begin{eqnarray}
R(D^\ast)_{\text{SM}} &=&  0.258 \pm 0.005,
\end{eqnarray}
The SM predictions for $R(D)$ from FLAG working group~\cite{Aoki:2016frl} is:
\begin{eqnarray}
R(D)_{\text{SM}} &=& 0.300 \pm 0.008\,
\end{eqnarray}
Refs.~\cite{Bernlochner:2017jka, Jaiswal:2017rve, Bigi:2017jbd} show that the SM error can be reduced to 0.003. The significance of $R(D)$ discrepancy does not change for these two values of SM error. We will quote results for both these cases in the results section.

In our model $W_R$ connects to both $\bar b_Rc_R$ current and the $\bar\tau_R\nu_{\tau,R}$ current leading to the effective operator:
\begin{eqnarray}
{\cal H}_{eff}\simeq \frac{g^2_R}{2M^2_{W_R}}\bar b_R\gamma_\mu c_R\bar{\nu}_{\tau_R}\gamma^\mu{\tau_R} + h.c.
\label{int1}
\end{eqnarray}
In the parity asymmetric model, we found that  the implication of the above flavor choice is that only the $b$ and $c$ quarks undergo quark seesaw.
The resulting $W_R$ interaction with quarks is given in Eq. (\ref{ccq}).
 Similarly in the lepton sector, only the tau-lepton field undergoes seesaw
which leads to lepton non-universal interaction of $W_R$ given in Eq. (\ref{ccl}), which helps us explain the $R(D^*,D)$ anomaly.

In the parity symmetric model, $W_R$ connection to the $\overline{b}_R \gamma_\mu c_R$ current arises from the Eq. (\ref{WR12}) while in the leptonic sector, only $\overline{\tau}_R \gamma_\mu \nu_{\tau_R}$ is allowed kinematically.  This is the case when $\nu_{\mu R}$ is heavier than 200 MeV or so. The $\nu_{eR}$ field couples to heavy leptons and $W_R$ and thus will not be relevant for $R(D^*,D)$ discussions.

\begin{figure}[h]
\centering
\includegraphics[height = 5cm]{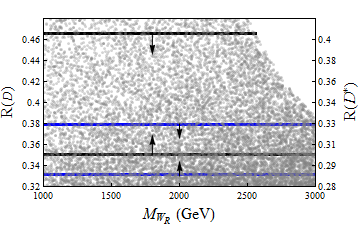}
\caption{$R(D,D^*)$ scatter-plot is shown by varying $g_R$ and $M_{W_R}$. The  boundaries of  $R(D)$ and $R(D^*)$ anomalies are shown by black and blue lines respectively. We show 1 $\sigma$ allowed regions.}
\label{allowed}
\end{figure}

\begin{figure}[h]
\centering
\includegraphics[height = 5cm]{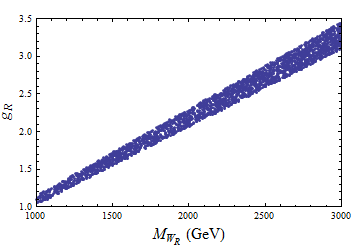}
\caption{$g_R$ vs $M_{W_R}$ in the allowed region of parameter space where  $R(D,D^*)$ anomalies are satisfied simultaneously.}
\label{coupling}
\end{figure}

To see if the interaction of Eq. (\ref{int1}) may explain the $R(D^*,D)$ anomaly,
we vary $g_R$ and $W_R$ and calculate $R(D^*,D)$. We show a scatter plot with points (in gray) which explains the anomaly in Fig.~\ref{allowed}. The allowed ranges of $R(D)$ and $R(D^*)$ anomalies are enclosed by the black lines and blue lines respectively. In Fig.~\ref{coupling} we show  $g_R$ as a function of $W_R$ mass in the  1 $\sigma$  allowed  overlapping regions (between top blue and bottom black curves) arising from the simultaneous explanations of  $R(D^*,D)$ anomalies. As can be seen in this figure, as $g_R$ increases $M_{W_R}$ takes larger values.

\section{Collider constraints: LHC and LEP}

Let us first focus on the constraints arising from a low mass $Z_R$ boson predicted by the model.
The coupling of $Z_R$ gauge boson to fermions is given by the Lagrangian (ignoring small $Z_L-Z_R$ mixing)
\begin{equation}
{\cal L}_{Z_R} = \frac{g_R^2}{\sqrt{g_R^2-g_Y^2}}\, \overline{f}_{L,R} \,\gamma_\mu \left[T_{3R} -\frac{Y_{L,R}}{2} \, \frac{g_Y^2}{g_R^2} \right]f_{L,R} \,Z_R^\mu~.
\label{int}
\end{equation}
Here $g_R$ is the $SU(2)_R$ gauge coupling, $g_Y$ is the hypercharge coupling given by $g_Y^2= 4 \pi \alpha/(1-s_W^2) = 0.1279$ (using values for the weak mixing angle $s_W^2(M_Z) = 0.2315$ and $\alpha(m_Z) = 1/127.9$).  $T_{3R} = \pm \frac{1}{2}$ or $0$ for $SU(2)_R$ doublets and singlets.  In the model under discussion, all left-handed fermions will have $T_{3R} = 0$.  $Y_{L,R}$ refer to the hypercharges of $f_{L,R}$ with the normalization $Y(e_R) = -2$. The $B-L$ gauge coupling $g_B$ appearing in the interactions has been replaced by the hypercharge coupling $g_Y$ using the formula that embeds $Y$ within $SU(2)_R \times U(1)_{B-L}$, see Eq. (\ref{embed}).
We shall treat $g_R$ as a variable parameter, but note that $g_R^2 \geq g_Y^2$ is required for consistency of Eq. (\ref{embed}). We shall demand that $g_R^2 \leq 4 \pi$ to stay within perturbative limits.

The decay width for $Z_R \rightarrow \overline{f} f$ to fermions of mass $m_f$ is given by
\begin{eqnarray}
\Gamma(Z_R \rightarrow \overline{f} f) = \frac{g_R^4}{g_R^2-g_Y^2}\, \frac{M_{Z_R}}{48 \pi}\,\beta \left[\frac{3-\beta^2}{2}\,a_f^2 + \beta^2\, b_f^2  \right] \label{width}
\end{eqnarray}
where
\begin{equation}
\beta = \sqrt{1-\frac{4 m_f^2}{M_{Z_R}^2}},~~~a_f = T_{3R} - \frac{Y_L+Y_R}{2}\, \frac{g_Y^2}{g_R^2},~~~b_f = T_{3R} - \frac{Y_R-Y_L}{2}\, \frac{g_Y^2}{g_R^2}~.
\end{equation}

In addition, $Z_R$ can decay into $W_L^+ W_L^-$ pair utilizing the small $Z_L-Z_R$ mixing and the SM $ZW^+ W^-$ vertex.  Although this partial decay width is suppressed by $\sin^2\xi$ ($\xi$ is the $Z_L-Z_R$ mixing angle), it is enhanced by a factor $(M_{Z_R}/M_{W_L})^4$, and could be significant.  The decay width is given by \cite{barger}
\begin{eqnarray}
\Gamma(Z_2 \rightarrow W^+ W^-) = \frac{g_L^2 \sin^2\xi}{192 \pi c_W^2} M_{Z_2} \left[\frac{M_{Z_2}}{M_{W}} \right]^4 \left[
1 - \frac{4 M_{W}^2}{M_{Z_2}^2} \right]^{3/2} \left[1 + 20 \frac{M_W^2}{M_{Z_2}^2} + 12 \frac{M_W^4}{M_{Z_2}^4}  \right]~.
\label{Z2WW}\end{eqnarray}
$Z_2$ can also decay into $h+ Z$.  The interaction Lagrangian for this decay in our model is given by
\begin{equation}
{\cal L}_{Z-Z_2-h} = g_Y^2 \sqrt{\frac{g_Y^2+g_L^2}{g_R^2-g_Y^2}}{1\over{\sqrt{2}}} \kappa_L Z_1^\mu Z_{2\mu} \,h \equiv f_{Z_1 Z_2 h} Z_1^\mu Z_{2\mu} h
\label{ZZ2h}\end{equation}
and the partial width is given by
\begin{equation}
\Gamma(Z_2 \rightarrow Z + h) = \frac{\left|f_{Z_1 Z_2 h}/M_{Z_1}\right|^2}{192 \pi} M_{Z_2} \lambda^{1/2}\left[1, \frac{M_{Z_1}^2}{M_{Z_2}^2},
\frac{M_{h}^2}{M_{Z_2}^2}\right]\left\{ \lambda \left[1, \frac{M_{Z_1}^2}{M_{Z_2}^2},
\frac{M_{h}^2}{M_{Z_2}^2}\right] + 12\frac{M_{Z_1}^2}{M_{Z_2}^2} \right\}~.  \\
\label{ZZ2h1}\end{equation}
Here $\lambda(a,b,c) \equiv a^2+b^2+c^2-2 ab-2 ac - 2bc$. In Eqs.~\ref{Z2WW}$\,-\,$\ref{ZZ2h1}, $Z_1$ can be identified as the SM  $Z$ and $Z_2$ as the heavy $Z_R$.

The branching ratios to various fermions follows from Eq. (\ref{width}).  Also the total width of $Z_R$ as a function of $g_R$ can be computed.  We consider two specific scenarios, one Parity asymmetric, and one Parity symmetric.

\subsection{Parity asymmetric scenario}

Here we focus on the case where all exotic fermions have masses larger than $M_{Z_{R}}/2$, so that $Z_R$ decays only into SM fermions and the three species of $\nu_R$, which are assumed to be light.  Furthermore, as discussed in Sec. 3, we shall assume a flipped scenario with respect to $SU(2)_R$, where the light chiral fermions $u_R$, $t_R$, $d_R$, $s_R$, $e_R$, $\mu_R$ are $SU(2)_R$ singlets (with $T_{3R} = 0$), while $c_R$, $b_R$, $\tau_R$ as well as the three flavors of $\nu_R$ belong to $SU(2)_R$ doublets with $T_{3R} = \pm1/2$.  Numerical values of the branching ratios defined as
\begin{eqnarray}
B_\ell &=& \frac{\Gamma(e^+e^-) + \Gamma(\mu^+ \mu^-)}{\Gamma_{\rm total}}, ~~~B_\tau = \frac{\Gamma(\tau^+ \tau^-)}{\Gamma_{\rm total}},~~~
B_\nu =  \frac{3\Gamma(\nu_L \bar{\nu}_L) + 3 \Gamma(\nu_R \bar{\nu}_R)}{\Gamma_{\rm total}}\nonumber \\
B_{\rm jet} &=& \frac{\Gamma(u\bar{u})+\Gamma(d\bar{d}) + \Gamma(s \bar{s}) + \Gamma(c\bar{c})+ \Gamma(b\bar{b})}{\Gamma_{\rm total}},~~~B_t=\frac{\Gamma(t \bar{t})}{\Gamma_{\rm total}}
\label{branchingexpression}\end{eqnarray}
as well as the total width over mass $(\Gamma_{\rm total}/M_{Z_R})$ for this scenario are presented for five different values of $g_R$ of interest in Table~\ref{tab:party-asym}. The BR  of $Z_R$ decaying to di-bosons   is  less than 1\% for the $R(D, D^\ast)$ allowed  parameter space.

\begin{table}[h!]
\centering
\begin{tabular}{ |c|c|c|c|c|c|c| }
 \hline
 $g_R$ & $B_\ell$ (\%) & $B_\tau$ (\%) & $B_\nu$ (\%) & $B_{\rm jet}$ (\%) & $B_t$ (\%) & $\frac{\Gamma_{\rm total}}{M_{Z_R}}$ (\%)\\
 \hline
1 & 1.89 & 6.6 & 35.4 & 54.98 & 1.07 & 3.3\\
 \hline
 1.5 & 0.349 & 8.55 & 32.6 & 58.25 & 0.20 & 7.3\\
 \hline
 2.0 & 0.11 & 9.2 & 31.5 & 59.11 & 0.061 & 13\\
 \hline
 2.5 & 0.043 &9.4 & 30.97 & 59.5 & 0.024 & 20.5\\
 \hline
 3.0 & 0.021 & 9.65 & 30.67 & 59.6 & 0.011 & 29.6\\
 \hline
\end{tabular}
\caption{Values of the branching ratios of $Z_R$ for decays into fermion pairs as a function of $g_R$ in the Parity asymmetric scenario.  $B_x$'s are defined in Eq.~(\ref{branchingexpression}). The last column lists the total width of $Z_R$ as a fraction of its mass.}
\label{tab:party-asym}\end{table}

As the value of $g_R$ increases, $B_\ell$ decreases dramatically, reaching $B_\ell = 1.1 \times 10^{-3}$ for $g_R = 2$.  This occurs due to the flipping of $e_R$ and $\mu_R$ with $E_{1R}$ and $E_{2R}$ under $SU(2)_R$ transformation, a feature facilitated by their common SM quantum numbers.  This flipping means that $e_R$ and $\mu_R$ carry zero $T_{3R}$ quantum number, and thus they interact with $Z_R$ with a coupling proportional to $g_Y^2/g_R$, see Eq. (\ref{int}).

Among the light fermions, $W_R$ couples to only $b_R$, $c_R$, $\tau_R$ and $\nu_{\tau R}$ with a coupling given by $g_R/\sqrt{2}$. The decay width of $W_R$ is found to be \begin{eqnarray}
{\Gamma_{\rm total}\over{M_{W_R}}} \{2.6\%, \, 6\%,\, 11\%,\,16.6\%,\,24\%\} ~~{\rm corresponding~ to} ~ g_R = (1, \,1.5,\, 2.0,\, 2.5,\, 3.0)~.
\label{W-width-asym}
\end{eqnarray}

\subsubsection{LEP constraints}

$e^+ e^-$ collision at LEP above the $Z$ boson mass provides significant constraints on contact interactions involving $e^+ e^-$ and any fermion pair.  As it turns out, in this Parity asymmetric scenario, the couplings of $Z_R$ with electron (as well as muon) are highly suppressed, and the LEP constraints are automatically satisfied for a TeV scale $Z_R$.  To see this, consider the effective Lagrangian involving $(e^+ e^-)$ and ($\mu^+ \mu^-)$ first, which can be read off from Eq. (\ref{int}):
\begin{equation}
{\cal L}_{\rm eff} =  - \frac{g_Y^4}{g_R^2-g_Y^2} \frac{1}{M_{Z_R}^2} \frac{1}{\{1+ (\Gamma_{\rm total}/M_{Z_R})^2\}^{1/2}}\left[\bar{e}_R \gamma_\mu e_R + \frac{1}{2}\bar{e}_L \gamma_\mu e_L\right] \left[\bar{\mu}_R \gamma^\mu \mu_R + \frac{1}{2}\bar{\mu}_L \gamma^\mu \mu_L\right]~.
\label{LEP1}
\end{equation}
While the larger $T_{3R}$ contribution is absent for $e^+ e^- \rightarrow e^+e^-\,,\,\mu^+ \mu^-$, it is present in the process $e^+ e^- \rightarrow \tau^+ \tau^-$ on the $\tau$ vertex.  The LEP constraint on the scale of contact interaction from this process is $\Lambda^-_{RR} > 8.7$ TeV.  This translates into a limit on $Z_R$ mass given by
\begin{eqnarray}
M_{Z_R} > \{573, \, 600,\, 607,\, 607,\, 603\}~{\rm GeV} ~~{\rm corresponding~ to} ~ g_R = (1, \,1.5,\, 2.0,\, 2.5,\, 3.0)~.
\label{LEP2}
\end{eqnarray}
We see that the constraints are rather weak, which are automatically satisfied with TeV scale $Z_R$. Since the $\mu^+\mu^-$  and $\tau^+\tau^-$ are not universal, we can not use the simultaneous $\mu,\,\tau$ fit limits. But even if we had used that, the constraints are weaker compared to the mass of $Z_R$ required to satisfy the $R(D,D^*)$ anomaly for a given $g_R$. Other processes such as $e^+ e^- \rightarrow \bar{c} c$ and $e^+ e^- \rightarrow \bar{b} b$ provide somewhat weaker constraints than the ones quoted in Eq. (\ref{LEP2}).

\subsubsection{LHC constraints}

Important constraints for this model arise from the resonant production of $Z_R$ and $W_R$ at the LHC.  Let us consider first the $Z_R$ production. Due to the flavor structure, the coupling of $Z_R$ with $u$ and $d$ quarks are suppressed (see Eq. (\ref{int})) with the couplings going as $g_Y^2/g_R$. With these suppression factors, the production cross-section of $Z_R$ at the LHC is  smaller compared with the $Z'$ associated with the sequential standard model. The cross-sections  are shown in Table \ref{LHC-asymmetric} using a K-factor $= 1.3$. The most dominant constraint arises from the dilepton (with $e$ and $\mu$) final states. The branching ratios are shown in Table \ref{tab:party-asym}. Combining the branching ratio with the production cross-section for the case of $g_R =1$, we find the cross-section to be  larger than the experimental constraint. However,  for $g_R=1.5$, $\sigma\times Br(Z_R\rightarrow l^+ l^-)$ is $2\times 10^{-4}$ pb where the experimental constraint is $<4\times 10^{-4}$ pb~\cite{Aaboud:2017buh,Sirunyan:2018exx} which is well satisfied. We found that the parameter space with $Z_R$ mass $>1.2$ TeV and $g_R>1.2$  is allowed by the current LHC constraint.
 The dijet resonance cross-section $\sigma\times Br(Z_R\rightarrow jj)$ is 0.29 pb (for $g_R = 1.5$) where the experimental cross-section is 0.6 pb~\cite{Aaboud:2017yvp,Sirunyan:2018xlo} and therefore $M_{Z_R}>1$ TeV with $g_R>1$ is allowed.

\begin{figure}[h]
\centering
\includegraphics[height = 5cm]{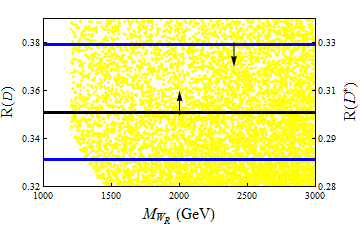}
\caption{LHC  allowed regions in the Parity asymmetric case.}
\label{allowedLHCsym}
\end{figure}

The  search for  $W_R$ is   difficult for this model, since it does not couple to the  first generation quarks. However, it can still be produced from the gluon-$b$ fusion and gluon-$c$ fusion as shown in Ref. \cite{Abdullah:2018ets}  since $W_R$ couples only to the $\overline{b}_R\gamma_\mu c_R$ current in the quark sector. In this case, the cross-section is suppressed compared to the case where $W_R$ couples to the $u,\,d$ partons in the protons. For example, for  $g_R=1$ and $W_R= 1$ TeV, the  $W_R$ production cross-section is 0.5 pb, which is allowed by the  direct search   ($\sim 0.6$ pb) with dijet final states~\cite{Aaboud:2017yvp,Sirunyan:2018xlo}. Similar conclusion holds for resonance searches with $\tau$$\nu$ final state~\cite{Aaboud:2018vgh,Sirunyan:2018lbg}. In Fig.~\ref{allowedLHCsym}, we show that  $M_{W_R}\geq 1.2$ TeV by the LHC  in the   $R(D, D^\ast)$  allowed region.

\begin{table}[h!]
\centering
\begin{tabular}{ |c|c|c|c|c|c|c| }
 \hline
 $g_R$ & $M_{Z_R}$(TeV)&$\sigma$(fb)\\
 \hline
 1.0 &1.0 & 0.8 \\\hline
 1.5 & 1.5 & $5.2\times 10^{-2}$ \\
 \hline
 2.0 & 2.0 & $7\times 10^{-3}$ \\
 \hline
 2.5 & 2.5 & $1.2\times10^{-3}$ \\
 \hline
 3.0 & 3.0 & $2.5\times10^{-4}$ \\
 \hline
\end{tabular}
\caption{$Z_R$ production cross-section at the LHC for the Parity asymmetric scenario }\label{LHC-asymmetric}
\end{table}

\subsection{Parity symmetric scenario:}

In this case we again assume all the exotic fermions have masses large enough to kinematically forbid $Z_R$ from decaying into those states.  $Z_R$ can then decay only into SM fermion pairs, as well as pairs of three $\nu_R$ species, which are assumed to be light.  In this case $(u_R, c_R)$ as well as ($d_R, b_R)$ are taken to be members of $SU(2)_R$ doublets with $T_{3R} = \pm 1/2$, as are ($\mu_R, \tau_R$) leptons.  On the other hand, $e_R$ belongs to $SU(2)_R$ singlet with $T_{3R} = 0$, a possibility which arises from the flipping of $e_R$ and $E_{1R}$.  Similarly, $t_R$ and $s_R$ are $SU(2)_R$ singlets. For this scenario, the branching ratios for $Z_R$ decays into various channels, as well as the total width to mass ratio of $Z_R$ are listed in Table~\ref{tab:parity-sym} as functions of $g_R$. The BR  of $Z_R$ decaying to di-bosons   is  less than 1\% for the $R(D, D^\ast)$ allowed  parameter space.

\begin{table}
\centering
\begin{tabular}{ |c|c|c|c|c|c|c| }
 \hline
 $g_R$ & $B_\ell$ (\%) & $B_\tau$ (\%) & $B_\nu$ (\%) & $B_{\rm jet}$ (\%) & $B_t$ (\%) & $\frac{\Gamma_{\rm total}}{M_{Z_R}}$ (\%)\\
 \hline
 1 & 3.6 & 3.2 & 16.9 & 64.82 & 11.5 & 5.9\\
 \hline
 1.5 & 3.89 & 3.82 & 14.58 & 65.26 & 12.42 & 14.12\\
 \hline
 2.0 & 4.08 & 4.05 & 13.87 & 65.27 & 12.71 & 25.7\\
 \hline
 2.5 & 4.17 & 4.16 & 13.56 & 65.26 & 12.83 & 40.61\\
 \hline
 3.0 & 4.22 & 4.22 & 13.41 & 65.25 & 12.90 & 58.84\\
 \hline
\end{tabular}
\caption{Values of the branching ratios of $Z_R$ for decays into fermion pairs as a function of $g_R$ in the Parity symmetric scenario. $B_x$'s are defined in Eq.~(\ref{branchingexpression}). The last column lists the total width of $Z_R$ as a fraction of its mass. }
\label{tab:parity-sym}\end{table}

As can be seen from Table~\ref{tab:parity-sym}, the branching ratio for $Z_R$ decaying into leptons is relatively stable under variations of $g_R$.  While $Z_R \rightarrow e^+e^-$ will drastically decrease with increasing $g_R$, the corresponding branching ratio for $Z_R \rightarrow \mu^+ \mu^-$ does not change much and contributes dominantly to $B_\ell$.  This has to do with the flipping of $e_R$ with $E_{1R}$, without flipping $\mu_R$ with $E_{2R}$ as was done in the Parity asymmetric scenario of Table~\ref{tab:party-asym}.  

Among the light fermions, $W_R$ couples to  $\overline{c}_R \gamma_\mu b_R$ as well as  $\overline{\mu}_R \gamma_\mu \nu_{\mu_R}$, and $\overline{\tau}_R \gamma_\mu \nu_{\tau_R}$.
The decay width of $W_R$ is found to be
\begin{eqnarray}
{\Gamma_{\rm total}\over{M_{W_R}}} \{3.3\%, \, 7.5\%,\, 13.3\%,\,20.7\%,\,29.8\%\} ~~{\rm for} ~ g_R = (1, \,1.5,\, 2.0,\, 2.5,\, 3.0)~.
\label{W-width-sym}
\end{eqnarray}

\subsubsection{LEP constraints}

In this scenario the LEP constraints are slightly stronger than those obtained in the case of Parity asymmetric scenario. However, the difference is not much.  Since the $Z_R$ couplings to $\mu$  and $\tau$ are the same,  we  use simultaneous $\mu,\,\tau$ fit limits which provides the strongest limit. LEP limit on the scale of contact interaction from this process $e^+ e^- \rightarrow l^+ l^-$ is $\Lambda^-_{RR} > 9.3$ TeV, which implies the following limits on $Z_R$ gauge boson mass:
\begin{eqnarray}
M_{Z_R} > \{611, \, 634,\, 638,\, 624,\, 598\}~{\rm GeV} ~~{\rm corresponding~ to} ~ g_R = (1, \,1.5,\, 2.0,\, 2.5,\, 3.0)~.
\end{eqnarray}
These constraints are slightly more stringent compared to the ones obtained from $e^+ e^- \rightarrow \tau^+ \tau^-$ in the Parity asymmetric scenario (see Eq. (\ref{LEP2})), but not by very much. All other LEP processes give weaker constraints.  We conclude that $Z_R$ mass of order 1 TeV is fully consistent with LEP data in this Parity symmetric scenario as well. It is to be noted that this weakened constraint is a result of the flipping of $e_R$ with $E_{1R}$.

\subsubsection{LHC constraints}

\begin{figure}[h]
\centering
\includegraphics[height = 5cm]{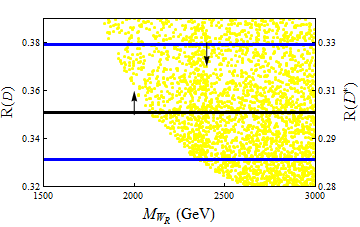}
\caption{LHC  allowed regions in the Parity symmetric case.}
\label{allowedLHC-sym}
\end{figure}

For the parity symmetric model, $Z_R$ (and $W_R$ for $V_R$ of the form  in Eq. (\ref{VR0})) are coupled to the first generation quarks with sizable couplings which make their production cross-sections large at the LHC. However due to large values of $g_R$, the model  has  large decay widths for $Z_R$ and $W_R$  for $g_R\geq 1$, see
Table \ref{tab:parity-sym} and Eq. (\ref{W-width-sym}). This causes problems in obtaining constraints at the LHC. The dilepton resonance search analyses which provide the best  constraint on the $Z_R$~\cite{Aaboud:2017buh,Sirunyan:2018exx} masses at the LHC are based on narrow resonances. In this final state, the maximum values of ${\Gamma\over{M_{Z_R}}}$ used in the analyses are  10\% for CMS~\cite{teruki} and $\sim$30\% for ATLAS~\cite{Aaboud:2017buh}.  For larger ${\Gamma\over{M_{Z_R}}}$, the constraint on the production cross-sections gets relaxed compared to the narrow resonance case, e.g., Ref. ~\cite{Aaboud:2017buh} shows that the cross-sections can be relaxed by a factor 2 for the maximum  $\Gamma/M_{Z_R}\sim 30$\% which  has been investigated. The dijet resonance search analysis also puts constraint on the $Z_R$ and $W_R$ masses~\cite{Aaboud:2017yvp,Sirunyan:2018xlo}, however, the LHC constraints exist for $\Gamma/{M_{Z_R, W_R}}\leq 30\%$.  The  constraint on the production cross-section gets relaxed  as $\Gamma/{M_{Z_R, W_R}}$ increases, e.g.,  for $\Gamma/{M_{Z_R, W_R}}\sim 30\%$, the constraint on the cross-section goes down by an order of magnitude~\cite{Sirunyan:2018xlo}. No LHC analysis exists for any final state where $\Gamma/M_{Z_R, W_R}> 30$\% which occurs when $g_R>2.2$. From Fig.~\ref{coupling} we see that $g_R>2.2$ can occur for $M_{W_R}>1.8$ TeV.  The larger width resonance is difficult to be extracted over a continuum background unless the experimental analysis would be able to reduce the background yield to a negligible level.  A new analysis is imperative to search for large decay width scenarios. In Fig.~\ref{allowedLHC-sym} we show the allowed region  of parameter space by the current LHC data. We see from here  that $M_{W_R}\geq1.8$ TeV is allowed.

\section{ Cosmological and astrophysical constraints}

In this section we comment on various cosmological and astrophysical constraints that should be satisfied by the model.  Some of the constraints arise from a light $\nu_{\tau R}$ needed for the $R(D^*,D)$ anomaly in our framework, while some others have to do with the adopted flavor structure.

\subsection{Supernova constraints}

A light $\nu_{R}$ may be produced inside the supernova core if its mass is below about 100 MeV.  This is indeed the case for $\nu_{\tau R}$ in our model for $R(D^*,D)$ anomaly.  If the interactions of the light $\nu_R$ with the supernova matter is too weak, the $\nu_R$ will escape, contradicting the observation of neutrino burst from SN1987a.  If the $\nu_R$ interacts with supernova matter it may be trapped inside, in which case the constraints will be relaxed.  Here we follow the crude analytic model studied in Ref. \cite{barbieri} to derive the allowed parameter space from SN1987a observations.

The model of Ref. \cite{barbieri} assumes a constant core density of $\rho_C \simeq 8 \times 10^{14}$ g/cm$^3$, corresponding to a total mass of $M \simeq 1.4 M_{\rm Sun}$ and a radius $R_C \simeq 10^6$ cm and a temperature $T_C = (30-70$ MeV.  For the calculation of right-handed neutrino sphere, the density profile outside the core was assumed to be $\rho(R) = \rho_C(R_C/R)^m$ with $m=3-7$.  This uncertainty in the density profile, as well as the uncertainty in the core temperature leads to considerable uncertainty in the $\nu_R$ interaction strength allowed by SN1987a observations.  Ref. \cite{barbieri} also assumes that the energy loss in $\nu_R$ emission should be less than about 20 times energy loss in $\nu_L$ emission.  Under these assumptions the following region in an effective mass $M_N$ was found to be excluded:
\begin{equation}
(2.4-4.3)M_{W_L} \leq M_N \leq (7.5-40) M_{W_L}~.
\end{equation}
This limit arises from the neutral current process $e^+ e^- \rightarrow \overline{\nu}_R \nu_R$, whose cross section was parametrized as
\begin{equation}
\sigma (e^+ e^- \rightarrow \overline{\nu}_R \nu_R) = \frac{G_F^2 s}{12 \pi} \left[\frac{M_{W_L}}{M_N} \right]^4~.
\end{equation}

In our model, the neutral current process $e^+ e^- \rightarrow \overline{\nu}_R \nu_R$ does occur.  The cross section for this process, both in the parity symmetric and asymmetric version, is given by
\begin{equation}
\sigma (e^+ e^- \rightarrow \overline{\nu}_R \nu_R) =\left(\frac{5}{16}\right) \frac{1}{48\pi} \frac{g_Y^4 g_R^4}{(g_R^2-g_Y^2)^2} \frac{s}{M^2_{Z_R}}~.
\end{equation}
The exclusion region is then obtained for various values of $g_R$ as
\begin{eqnarray}
&~& (239-429)~ {\rm GeV} \leq M_{Z_R} \leq (748-3890)~ {\rm GeV}~~~~(g_R = 2) \nonumber \\
&~& (252-452) ~{\rm GeV} \leq M_{Z_R} \leq (788-4203)~{\rm GeV}~~~~(g_R = 1)~.
\end{eqnarray}
In these exclusion regions, one should take the weaker limit, which is found to be consistent with the range of parameters needed for explaining the $R(D^*,D)$ anomaly. We note in passing that the charged current $W_R$ interactions does not lead to neutronization process $e_R p \rightarrow \nu_R n$ in our model, since $W_R^\pm$ has no coupling to the electrons.

\subsection{Other constraints}

In the parity asymmetric model,
in the heavy quark sector, we see that in the limit of $\epsilon_i\to 0$ in Eq. (\ref{VRa}), the lightest of the $(U,D, S, T)$ quarks will remain stable and will not annihilate fast enough so that it can over-close the universe. The reason for this is that for TeV mass colored particles, the only annihilation channel for $T\leq M_Q$, is via gluon emission to two light quarks i.e. $Q\bar{Q}\to q \bar{q}$.  This cross section goes as $\sigma_{Q\bar{Q}}\sim \frac{\alpha^2_s}{M^2_Q}$ which is $\leq {\rm pb}$. This implies that they could either form the dark matter of the universe, which is unacceptable since these are colored particles or worse, they over-close the universe. We therefore need for the lightest of the heavy quarks to decay. Once $\epsilon_{1,2}$ are turned on in our model, the relevant lightest heavy quark can decay to $b$ and $c$ quarks which decay via the left-handed CKM matrix $V_L$  to leptons and follow the usual cosmology. Typical decay rate for these fermions can be estimated to be $\Gamma_Q\sim \frac{g^4_R}{192\pi^3 M^4_{W_R}}\epsilon^2M^5_Q$ and the temperature at which they will decay can be estimated by using
\begin{eqnarray}
\Gamma_Q\sim g^{1/2}_*\frac{T^2_d}{M^2_{Pl}}~.
\end{eqnarray} For these decays to happen above a Temperature of the universe $T > 1$ GeV, we need $\epsilon_{1,2}\geq 10^{-9}$ \cite{DMZ}. This is a rather weak constraint and is therefore easily satisfied without contradicting any other phenomenology.
Similarly, in the lepton sector, we can introduce small mixings among the right handed leptons to make the heavy neutral and charged leptons to decay above $T\sim  1$  GeV to avoid conflict with BBN requirements.

The existence of light $\nu_R$ states can modify big bang nucleosynthesis.  If the $\nu_R$ decouples from the plasma above QCD phase transition temperature, then their contribution to effective neutrino species is about 0.1 per $\nu_R$ species.  Even with all three neutrinos being light, this excess is consistent with BBN constraints.  As noted in footnote 4, if the light $\nu_R$ also play a role in short baseline neutrino anomalies from LSND and MiniBoone, then large scale structure formation constraints become important \cite{Planck} within the $\Lambda$CDM paradigm.  Secret neutrino interactions can potentially relax these limits \cite{br}. We have not explored this possibility here.

\section{ Discussion and conclusion}

Before concluding, we make a few observations of theoretical nature on the model presented here.

{\bf 1.} In the parity asymmetric model, we have several vector-like fermions acquiring masses from the right-handed Higgs mechanism.  As seen from Eq. (\ref{heavy}) and Eq. (\ref{heavylep}), the masses of $U,T, D,S$ quarks as well as $E_1$ and $E_2$ leptons arise from $Y_i \kappa_R$. Perturbativity of the Yukawa couplings would then imply that these vector-like fermions have masses not much above $\kappa_R \simeq (1.5-2.5)$ TeV.  This can be made more precise by looking at partial wave unitarity in the process $\overline{f} f \rightarrow \overline{f} f$ mediated by the $Z_R$ and $W_R$ gauge bosons.  Such an analysis in the context of the SM leads to a limit of 550 GeV on the mass of a fourth generation quark \cite{chanowitz}. For $N$ generations of quarks, this is strengthened by a factor of $1/\sqrt{N}$.  These results can be readily scaled up to the masses of vector-like quarks of our model.  We find for four degenerate quarks, $M_Q \leq 2.24$ TeV, for $M_{W_R} = 2$ TeV and $g_R = 2.0$. Other processes, such as $\overline{f} f \rightarrow W_R^+ W_R^-$ can also yield useful limits.  Using the results of Ref. \cite{bjz} we obtain $M_Q \leq 5.6$ TeV, which is somewhat weaker. In the parity symmetric scenario the mass of the vector like partner of electron is given as $Y \kappa_R$.  The partial wave unitarity limit on a fourth generation SM lepton mass is 1 TeV, which can be scaled to obtain a limit of $M_E \leq 4.5$ TeV for the vector-like partner of the electron. In the $P$ asymmetric case, since two such vector-like leptons acquire their masses from $\kappa_R$, the partial wave unitarity limit on their (common) mass is $M_E < 3.2$ TeV.

{\bf 2.} The boundedness of the Higgs potential of Eq. (\ref{V}) poses an upper limit on the masses of fermions generated by the Higgs mechanism.  In the parity asymmetric model, four quarks and two leptons acquire such masses.  The quartic coupling $\lambda_{1R}$ will turn negative at higher energies if these Yukawa couplings are large.  This should not happen at least for an order of magnitude higher energy.  Demanding this would lead to an upper limit on vector-fermion masses.  To see this, we can examine the renormalization group evolution equation for $\lambda_{1R}$, which is given by
\begin{eqnarray}
&~&16 \pi^2 \frac{d \lambda_{1R}}{dt} = 12 \lambda_{1R}^2 + 4 \lambda_2^2- \lambda_{1R}(3 g_B^2 + 9g_R^2) +\frac{3}{4} g_B^4+ \frac{3}{2} g_B^2 g_R^2 + \frac{9}{4} g_R^4 + \nonumber \\
&~& \lambda_{1R} {\rm Tr}\left(3 Y_U^{\prime \dagger} Y_U' + 3 Y_D^{\prime \dagger} Y_D'+  Y_E^{\prime \dagger} Y_E'\right)
-4 {\rm Tr}\left(3 (Y_U^{\prime \dagger} Y_U')^2 + 3 (Y_D^{\prime \dagger} Y_U')^2+  (Y_E^{\prime \dagger} Y_E')^2 \right)~.
\end{eqnarray}
The full set of RGE for the Yukawa couplings in a closely related universal seesaw model can be found in Ref. \cite{bgk}.
With four degenerate quark  and two lepton fields, demanding that $\lambda_{1R}$ remains positive up to a scale of $10 \kappa_R$ gives a limit on these fermion masses of about 2.5 TeV.  This limit depends on the initial value of $\lambda_{1R}$. The upper limit on vector-like fermion masses are $M_F \leq (1.6, \,1.9, \,2.2,\, 2.5)$ TeV, corresponding to the initial value of $\lambda_{1R} = (1.0, \,2.0,\, 3.0,\, 4.0)$ (keeping $\lambda_2$ fixed at $0.7$).

{\bf 3.} We have used relatively large values of the $SU(2)_R$ gauge coupling $g_R$.  However, perturbation theory is still valid, as the theory is asymptotically free.  If the Higgs fields of the model are not present, the $SU(2)_R$ theory is one with $N_f = 6$ (that is, with twelve doublets), which has been studied non-perturbatively on the lattice \cite{lattice1,lattice2}.  The phase diagram of such a theory appears to be emerging, with the $N_f = 6$ lying close to the boundary of the conformal window.  Since we Higgs the theory, the gauge coupling $g_R$ increases coming from higher to lower energies, until the Higgsing occurs.  A fixed point value of $g_*^2 \simeq 14.5$ was found in Ref. \cite{lattice2}.  Just before the theory acquires this fixed point value, we assume that spontaneous symmetry breaking occurs.  A semi-perturbative value of $g_R \sim (2.0-3.0)$ appears quite reasonable in this case.

{\bf 4.} Our model (and the general universal seesaw models) does not grand unify into conventional GUT groups such as $SU(5)$ or $SO(10)$.  However, these models can be embedded into grand unified symmetries based on $SU(5) \times SU(5)$ or $SO(10) \times SO(10)$.  For the former possibility and as one example how unification works in such models, see Ref. \cite{LM}. The unification of gauge couplings occurs in multiple steps, and therefore is a bit nontrivial.  Proton decay mediated by the gauge bosons in such models leads to the dominance of $p \rightarrow e^+ \pi^0$ decay mode with a lifetime estimated to be near the current experimental limit.

In summary, we have presented a UV complete theory that resolves the $R(D^*, D)$ anomaly based on left-right gauge symmetry with a low mass $W_R$ and a relatively large $g_R$. Two versions of the theory were developed, one with softly broken parity symmetry and one without parity.  In the former case the model solves the strong CP problem with parity symmetry, without invoking the Peccei-Quinn symmetry and the resulting axion.  In each case we have presented flavor structures that lead to a consistent explanation of the $R(D^*,D)$ anomaly in terms of the right-handed currents, which are also compatible with low energy flavor violation constraints.  The charged $W_R^\pm$ that mediates new contributions in $B$ decays is accompanied by a neutral $Z_R^0$, which is nearly degenerate in mass with the $W_R^\pm$.  LEP and LHC experiments provide stringent limits on these relative low mass gauge bosons.  Their discovery would be somewhat challenging, since their total widths turn out to be 20\% or more compared to their masses, once the $R(D^*,D)$ anomaly is explained.

The parity asymmetric version of the model has several vector-like quarks that acquire masses via the Higgs mechanism. These masses cannot be greater than about 2.5 TeV, to be consistent with perturbative unitarity and an understanding of $R(D^*,D)$ anomaly.  In the parity symmetric version, the top quark partner is predicted to have a mass $M_T = (1.5-2.5)$ TeV.  A vector-like electron partner with a mass less than 4.5 TeV is expected in both cases.  Along with the gauge bosons, these vector-like fermions provide a rich spectrum waiting to be explored at the LHC.

\vskip0.2in

\section*{Acknowledgements} We would like to thank Julian Calle, Bogdan Dobrescu, Ricardo Eusebi, George Fleming, Sudip Jana, Teruki Kamon and Jure Zupan for very helpful discussions. The work of KSB is supported by U.S. Department of Energy Grant No. de-sc0016013 and by a Fermilab Distinguished Scholar program. The work of BD is supported by DOE Grant No. de-sc0010813. The work of RNM is supported by the US National Science Foundation under Grant No. PHY1620074. KSB is thankful to the Mitchell Institute at Texas A\& M University for hospitality during the workshop on ``Collider, Dark Matter and Neutrino Physics, 2018". KSB and RNM acknowledge hospitality of the Bethe Center for Theoretical Physics, University of Bonn during the workshop on ``Grand Unification and the Real World". KSB and BD are thankful to the Theory Group at Fermilab for hospitality during a summer visit.

\end{document}